\documentclass[a4paper]{elsarticle}
\usepackage{amssymb,amsmath,amsthm,graphicx,hyperref,tabularx}
\usepackage[top=2.5cm,bottom=2.5cm,left=2.5cm,right=2.5cm]{geometry}
\usepackage{booktabs}
\usepackage{tikz}
\usetikzlibrary{arrows.meta,positioning,fit}
\graphicspath{{Figs/}}
\usepackage{subcaption}
\begin{document}
\begin{frontmatter}

\title{A Systematic Review and Taxonomy of Reinforcement Learning-Model Predictive Control Integration for Linear Systems}
\author[1]{Mohsen Jalaeian Farimani \corref{cor1}}
\ead{mohsen.jalaeian@polimi.it}
\author[2]{Roya Khalili Amirabadi}\ead{roya.khalili.a@gmail.com}
\author[3]{Davoud Nikkhouy}\ead{davoudnikkhoy@gmail.com}
\author[3]{Malihe Abdolbaghi}\ead{ma.abdolbaghi@gmail.com}
\author[3]{Mahshad Rastegarmoghaddam}\ead{mahshad.rastegarmoghaddam@mail.polimi.it}
\author[3]{Shima Samadzadeh}\ead{shima.samadzade@gmail.com}
\author[2]{Mahdi Ghane}\ead{mahdi.ghane@hvl.no}
\cortext[cor1]{Corresponding author}
\affiliation[1]{organization={Department of Electronics, Information and Bioengineering (DEIB), Politecnico di Milano}, city={Milan}, country={Italy}}
\affiliation[2]{organization={Department of Computer science, Electrical engineering and Mathematical sciences, Western Norway University of Applied Sciences}, city={Haugesund}, country={Norway}}
\affiliation[3]{organization={Department of Mechanical Engineering, Politecnico di Milano}, city={Milan}, country={Italy}}
\begin{abstract}
The integration of Model Predictive Control (MPC) and Reinforcement Learning (RL) has emerged as a promising paradigm for constrained decision-making and adaptive control. MPC offers structured optimization, explicit constraint handling, and established stability tools, whereas RL provides data-driven adaptation and performance improvement in the presence of uncertainty and model mismatch. Despite the rapid growth of research on RL--MPC integration, the literature remains fragmented, particularly for control architectures built on linear or linearized predictive models. This paper presents a comprehensive Systematic Literature Review (SLR) of RL--MPC integrations for linear and linearized systems, covering peer-reviewed and formally indexed studies published until 2025. The reviewed studies are organized through a multi-dimensional taxonomy covering RL functional roles, RL algorithm classes, MPC formulations, cost-function structures, and application domains. In addition, a cross-dimensional synthesis is conducted to identify recurring design patterns and reported associations among these dimensions within the reviewed corpus. The review highlights methodological trends, commonly adopted integration strategies, and recurring practical challenges, including computational burden, sample efficiency, robustness, and closed-loop guarantees. The resulting synthesis provides a structured reference for researchers and practitioners seeking to design or analyze RL--MPC architectures based on linear or linearized predictive control formulations.
\end{abstract}

\begin{keyword}
Model Predictive Control (MPC), Reinforcement Learning (RL), Hybrid Control Architecture, Taxonomies.
 \end{keyword} 

\end{frontmatter}

\section{Introduction}
\label{sec:introduction}

MPC is widely recognized as one of the most successful advanced control strategies for constrained multivariable systems in both academia and industry. Its prominence is mainly due to its ability to explicitly account for system dynamics, operational constraints on states and inputs, and performance optimization over a finite prediction horizon in a receding-horizon framework \cite{QIN2003733, MAYNE20142967, Fundamental_Rawlings2017}. Owing to these advantages, MPC has been extensively adopted in process industries, energy systems, automotive control, and autonomous systems \cite{camacho2007mpc}.

At the core of MPC lies the use of an internal prediction model to forecast future system behavior and compute the control action by repeatedly solving an optimization problem online. This model-based nature is a major strength, but also a critical limitation when accurate first-principles models are difficult to obtain or maintain. In practical applications, nonlinearities, parametric uncertainty, time-varying dynamics, and unmodeled disturbances may substantially deteriorate controller performance if the prediction model does not adequately represent the real process \cite{MAYNE20142967, Fundamental_Rawlings2017}. In addition, the repeated online solution of constrained finite-horizon optimization problems may impose a significant computational burden, particularly in fast systems or embedded implementations with limited resources \cite{Wang2009, Fundamental_Rawlings2017}.

In contrast, RL has emerged as a powerful framework for sequential decision-making under uncertainty, enabling agents to learn control policies directly through interaction with the environment \cite{SuttonBarto2018,KHALILIAMIRABADI2026465}. Unlike conventional model-based control approaches, RL can improve performance from data and experience, which makes it especially appealing for systems with incomplete models or complex dynamics. Recent progress in deep reinforcement learning has further demonstrated the capability of RL to address high-dimensional and nonlinear decision problems that are difficult to solve using classical approaches alone \cite{Lillicrap2016, Mnih2015}.

Despite these advances, the direct use of RL in safety-critical control applications remains challenging. Standard RL algorithms generally do not guarantee recursive feasibility, closed-loop stability, or strict satisfaction of state and input constraints during either training or deployment. Moreover, pure trial-and-error learning may lead to unsafe exploration, which is unacceptable in many engineering systems \cite{GarciaFernandez2015, Brunke2022SafeLearningRobotics,Roya_HPV}. These issues have motivated increasing interest in safe and constrained learning-based control frameworks that can combine the adaptability of RL with the reliability of model-based control \cite{Brunke2022SafeLearningRobotics}.

Against this background, the integration of RL and MPC has attracted substantial attention as a promising hybrid control paradigm. The rationale is that MPC can provide structure, safety, and constraint handling, while RL can introduce adaptation, performance improvement, and compensation for model inaccuracies through data-driven learning. Existing studies have explored several forms of integration, including RL-based tuning of MPC parameters, learning of terminal costs or value functions, approximation of control laws, and adaptation of internal models or cost functions within the MPC loop \cite{KiumarsiEtAl2018, HewingKabzanZeilinger2020, GrosZanon2020}.

However, despite the rapid expansion of research at the intersection of RL and MPC, the literature remains fragmented across application domains, algorithmic choices, and implementation strategies. Many published studies address specific case studies or particular methodological aspects, while comparatively few works provide a structured and comprehensive synthesis of how RL and MPC are combined, why certain integrations are preferred, and under which problem settings they are most effective. This fragmentation creates a barrier to systematic understanding and makes it difficult for researchers and practitioners to identify consistent design patterns, methodological trade-offs, and open research directions.

To address this gap, this paper presents a SLR on the integration of RL and MPC for dynamic control systems. Following established review principles for evidence-based synthesis in engineering research \cite{kitchenham2007guidelines, xiao2020guidelines, okoli2015guide}, the study systematically identifies, screens, classifies, and analyzes the relevant body of literature in order to provide a coherent overview of this rapidly evolving field.

The main contributions of this paper are as follows:
\begin{itemize}
\item \textbf{Systematic methodological synthesis:} A rigorous and reproducible SLR of MPC--RL integration is conducted, using established systematic review procedures.
        
\item \textbf{Multi-dimensional taxonomy:} The existing literature is organized into a multi-dimensional taxonomy that captures the role of RL within the MPC framework,  the class of RL algorithms employed,  the adopted MPC formulation, and  the structure of the underlying cost function.
        
\item \textbf{Cross-dimensional analysis:} Cross-cutting relationships among controller architecture, learning strategy, control objective, and application domain are systematically analyzed to reveal recurrent design patterns and emerging trends. 

\item \textbf{Research gaps and future directions:} Open challenges are identified -- including safe learning, computational complexity, sample efficiency, and theoretical guarantees -- and a set of promising directions for future research on MPC--RL integration in linear and linearized control systems is outlined.
\end{itemize}

\textbf{Scope: }This review focuses on MPC-RL integrations in which the linear MPC is the sole primary controller-the control signal is determined entirely by solving an optimization problem based on a linear or linearized model. The RL agent serves exclusively as a tuner of MPC-internal parameters, such as weighting matrices, prediction horizons, terminal costs, or soft-constraint penalties. RL is not used for model learning, reference generation, nor as a supervisor that provides an additional control signal; it neither replaces the predictive model nor alters the MPC optimization structure. This tuning-centric paradigm is treated as the foundational integration strategy. Architectures where MPC acts as an expert actor, critic, or safety filter are outside this scope and are referred to existing literature (\cite{REITER2026101045}). The paper provides a systematic taxonomy of tuning-oriented RL-MPC combinations specifically for linear and linearized systems.

The remainder of this paper is organized as follows. Section 2 outlines the overall research approach adopted in this study. Section 3 describes the systematic review protocol, including the search strategy, study selection procedure, and data extraction methodology. Section 4 provides the theoretical background on MPC and RL. The proposed taxonomies are then presented and discussed in Sections 5-8. Section 9 reviews the principal application domains identified in the literature, followed by a cross-dimensional analysis in Section 10. Finally, Section 11 concludes the paper and highlights key findings, research limitations, and directions for future work.

\section{Research Approach} 
This research employs a SLR methodology to comprehensively analyze and synthesize existing research on the integration of MPC and RL for linear and linearized control systems. The SLR methodically identifies, evaluates, and synthesizes the literature to develop structured taxonomies and identify patterns, trends, and research gaps in MPC-RL hybrid approaches.

\subsection{Problem Statement}
The integration of MPC and RL has emerged as a promising paradigm for enhancing control system performance, adaptability, and robustness in linear and linearized control applications \cite{Gros2022_}. However, the field of MPC-RL integrations lacks a unified framework and systematic synthesis of its rapidly expanding literature. This fragmented knowledge makes it challenging for researchers and practitioners to navigate the diverse approaches, select appropriate techniques for specific applications, and understand the fundamental trade-offs involved. Furthermore, the absence of comprehensive taxonomies and comparative analyses hinders the identification of dominant trends, methodological patterns, and promising research directions. Consequently, in the domain of MPC--RL integrations for linear and linearized control systems, the following research challenges have been identified:

\begin{itemize}
\item \textbf{Fragmented Knowledge and Lack of Synthesis:} 
Concepts, methodologies, and algorithms for combining MPC and RL are scattered across extensive literature spanning control theory, machine learning, and various application domains. Creating a coherent taxonomy that captures the diverse roles of RL in MPC frameworks, the types of RL algorithms employed, the variations in MPC formulations, and their applications presents a significant challenge. A structured overview is essential for informed method selection and understanding of inter-method relationships.

\item \textbf{Understanding Trade-offs and Integration Strategies:} 
Different MPC--RL integration strategies embody distinct trade-offs across computational complexity, performance guarantees, adaptability, and robustness \cite{He202519266_}. The choice of RL role (planner, balancer, architect, observer, or guardian) involves specific trade-offs that are rarely analyzed systematically across different linear and linearized control applications and system characteristics.

\item \textbf{Identifying Trends and Emerging Patterns:} 
The rapid evolution of MPC--RL methods makes it challenging to identify dominant trends, stagnating approaches, and promising research directions without a longitudinal and comparative analysis of the field. The relationship between RL algorithm types, their assigned roles in MPC frameworks, and the resulting control performance requires systematic investigation.

\item \textbf{Assessment and Evaluation Challenges:} 
Performance assessments of MPC--RL methods are highly sensitive to experimental design, choice of benchmarks, baseline implementations, and evaluation metrics. The diversity in cost function formulations, constraint handling methods, and MPC types complicates fair comparison and reproducibility across studies.

\item \textbf{Computational and Implementation Concerns:} 
Many MPC--RL integrations incur substantial computational complexity, particularly when RL is used for real-time adaptation of MPC parameters or structures. Understanding the scalability and practical implementation constraints of different integration strategies is crucial for real-world deployment.

\item \textbf{Decision-Making Framework:} 
There is no comprehensive framework that integrates MPC--RL taxonomies, performance trade-offs, computational considerations, and application-specific guidelines to support principled method selection and deployment in linear and linearized control systems.
\end{itemize}

These challenges motivate the need for a structured methodology that systematically maps MPC--RL methods, identifies methodological patterns, and evaluates their implications for practical control system design and implementation.

\subsection{Research Questions}
To guide our systematic inquiry into MPC--RL integrations for linear and linearized control systems, we formulated the following Research Questions (RQs):

\begin{description}
\item[\textbf{RQ1:}] What are the specific limitations of classical Linear MPC that motivate the use of RL, and what specific performance metrics (e.g., computational time, adaptability, optimality gap) are improved through this integration?

\item[\textbf{RQ2:}] What are the structural patterns and dominant architectures for integrating RL into Linear MPC, and how do they differ in terms of the RL agent’s role (e.g., tuner, approximator, or supervisor)?

\item[\textbf{RQ3:}] What are the relationships and co-occurrences between different taxonomic categories (e.g., between RL roles and RL algorithm types, between MPC formulations and application domains, between cost functions and RL roles) in MPC--RL integrations?

\item[\textbf{RQ4:}] What temporal trends and patterns emerge in the development and application of MPC--RL integrations, and how have research priorities evolved across different dimensions of the taxonomies?

\item[\textbf{RQ5:}] How do existing studies address the critical challenges of stability guarantees, constraint satisfaction, and sample efficiency when applying RL to linear and linearized control systems?

\item[\textbf{RQ6:}] What practical insights and guidelines can be derived from the synthesized findings to inform the selection, design, and implementation of MPC--RL integrations for specific linear and linearized control applications?

\item[\textbf{RQ7:}] Based on the synthesis of current methods, what guidelines can be established for selecting the appropriate MPC--RL architecture for specific linear applications, and what are the open research gaps?
\end{description}

\subsection{Research Methods}
This study adopts a SLR methodology to comprehensively identify, analyze, and synthesize existing research on MPC--RL integrations for linear and linearized control systems. The review was conducted in accordance with established SLR guidelines \cite{kitchenham2007guidelines}, ensuring methodological rigor, transparency, and reproducibility.

The SLR process was designed to systematically address the research questions outlined in Section 2.2 and consisted of the following key stages:

\begin{enumerate}
\item \textbf{Review Planning:} 
A review protocol was defined prior to the execution of the study. This stage involved the formal specification of the research objectives, formulation of the research questions, and identification of key methodological decisions governing the review process.

\item \textbf{Search Strategy:} 
A structured and comprehensive search was conducted across major academic digital libraries. The search strategy employed carefully constructed search strings comprising terms related to ``Model Predictive Control'', ``Reinforcement Learning'', ``MPC--RL'', and their variants, with the aim of maximizing coverage while maintaining relevance.

\item \textbf{Study Selection:} 
Retrieved studies were screened using predefined inclusion and exclusion criteria to ensure relevance and quality. The selection procedure followed a two-stage process consisting of (i) title and abstract screening and (ii) full-text assessment. The entire screening process was documented to ensure transparency and replicability.

\item \textbf{Data Extraction and Synthesis:} 
For each included study, relevant information was extracted using a standardized data extraction form. A thematic synthesis was then performed to identify recurring patterns and relationships, culminating in the development of the proposed taxonomies and the comparative analyses presented in this review.
\end{enumerate}

\section{Systematic Literature Review Methodology}
\label{sec:methodology}

This study follows a rigorous SLR methodology in accordance with established guidelines proposed in \cite{kitchenham2007guidelines}, \cite{xiao2020guidelines}, and \cite{okoli2015guide}. The objective of the SLR is to systematically identify, evaluate, and synthesize existing research on MPC--RL integrations for linear and linearized control systems.

\subsection{Review Protocol}

The SLR was conducted through the following structured stages:

\begin{enumerate}
\item \textbf{Problem Definition and Research Questions:} 
The review process began with the formal definition of the problem scope and formulation of the research questions, ensuring that the SLR addressed focused and answerable objectives.

\item \textbf{Search Strategy and Data Sources:} 
A comprehensive search strategy was developed based on an analysis of seminal MPC and RL studies and commonly used terminology. The resulting search strings were applied to major academic digital libraries, including IEEE Xplore, ACM Digital Library, Scopus, Web of Science, and Springer Link, targeting publication metadata (title, abstract, and keywords). The review focuses on peer-reviewed and formally indexed publications until 2025.
 Our primary search strings were as follows:
\textit{("Model Predictive Control" OR "MPC") AND ("Reinforcement Learning" OR "RL" OR "Deep Reinforcement Learning" OR "DRL") AND ("linear system" OR "linear control" OR "LTI" OR "linear time-invariant")}, and 
\textit{("MPC--RL" OR "RL--MPC" OR "reinforcement learning model predictive control") AND ("linear" OR "control" OR "system")}, and
\textit{("adaptive MPC" OR "learning-based MPC" OR "RL for MPC" OR "MPC tuning" OR "MPC adaptation") AND ("reinforcement learning")}

\item \textbf{Study Selection and Screening:} 
Retrieved publications were merged and deduplicated. A two-stage screening process--consisting of title/abstract screening followed by full-text assessment--was performed using predefined inclusion and exclusion criteria. The screening process was documented using a PRISMA-style flow diagram to ensure transparency.

\textbf{Inclusion Criteria:} Papers were included in the review if they satisfied the following conditions: (IC1) the study proposes, implements, or analyzes an integration of MPC and RL for linear control systems; (IC2) the focus is on linear systems or linearized system models; (IC3) the paper is peer‑reviewed or formally indexed; (IC4) the publication is written in English; (IC5) the full text of the paper is accessible; and (IC6) the study provides sufficient methodological detail to enable classification within the proposed taxonomies.

\textbf{Exclusion Criteria:} Studies were excluded if any of the following conditions applied: (EC1) the work corresponds to short papers (less than six pages), posters, tutorials, or editorials without substantial methodological contribution; (EC2) the paper is a survey or tertiary review, although such works were used for snowballing and contextual understanding; (EC3) the study focuses exclusively on nonlinear systems without linear formulations or approximations; (EC4) the paper represents a duplicate or derivative version of an already included publication; or (EC5) Model Predictive Control or Reinforcement Learning is mentioned only peripherally without substantial methodological integration.

\item \textbf{Quality Assessment:} 
The methodological quality of the selected studies was evaluated using a set of predefined quality criteria adapted from \cite{okoli2015guide}, focusing on clarity of contributions, adequacy of methodological descriptions, soundness of experimental evaluation, and relevance to linear control systems. Only studies meeting the quality threshold were included in the final synthesis.

\item \textbf{Data Extraction and Synthesis:} 
Key information was extracted using a standardized form including: (1) publication details, (2) RL algorithm type and role, (3) MPC formulation characteristics, (4) cost function structure, (5) application domain, (6) experimental setup and results, and (7) reported strengths and limitations. A thematic synthesis was then conducted to construct the proposed taxonomies and to answer the research questions.
\end{enumerate}

To ensure completeness, both backward and forward snowballing were applied to the final set of included studies. The resulting curated collection of 60 publications constitutes a comprehensive knowledge base underpinning the taxonomies and analyses presented in this review.

\subsection{Search Process Execution}
\label{subsec:search_process}

The search and study selection were carried out through a structured four-phase process. This multi-stage procedure ensured a transparent, systematic, and reproducible transition from the initial retrieval of studies to the final corpus used for synthesis.

\begin{description}
\item[Phase 1: Pool of Publications] 
Automated searches across major digital libraries, complemented by manual searches, resulted in an initial pool of approximately 2,500 potentially relevant publications.

\item[Phase 2: Publication Pruning] 
The initial pool was refined by applying predefined inclusion and exclusion criteria through title and abstract screening, reducing the number of candidate studies to 180 publications that aligned with the research objectives.

\item[Phase 3: Quality Assessment] 
A rigorous quality assessment was conducted through full-text review to evaluate the methodological soundness and relevance of the selected studies. This phase resulted in 70 high-quality publications that satisfied the defined quality threshold and were retained for further analysis.

\item[Phase 4: Data Extraction and Taxonomy Development] 
Comprehensive data extraction was performed on the 60 retained studies. The extracted data was analyzed to develop the five-dimensional taxonomy (RL roles, RL algorithm types, cost function types, MPC types, and application domains) and to identify relationships between different taxonomic categories.
\end{description}

Overall, this carefully designed selection process ensured that the final dataset was comprehensive, methodologically robust, and well-suited to support the taxonomy development and comparative analyses presented in this review.

\begin{figure}[h!]
\centering
\includegraphics[width=\textwidth]{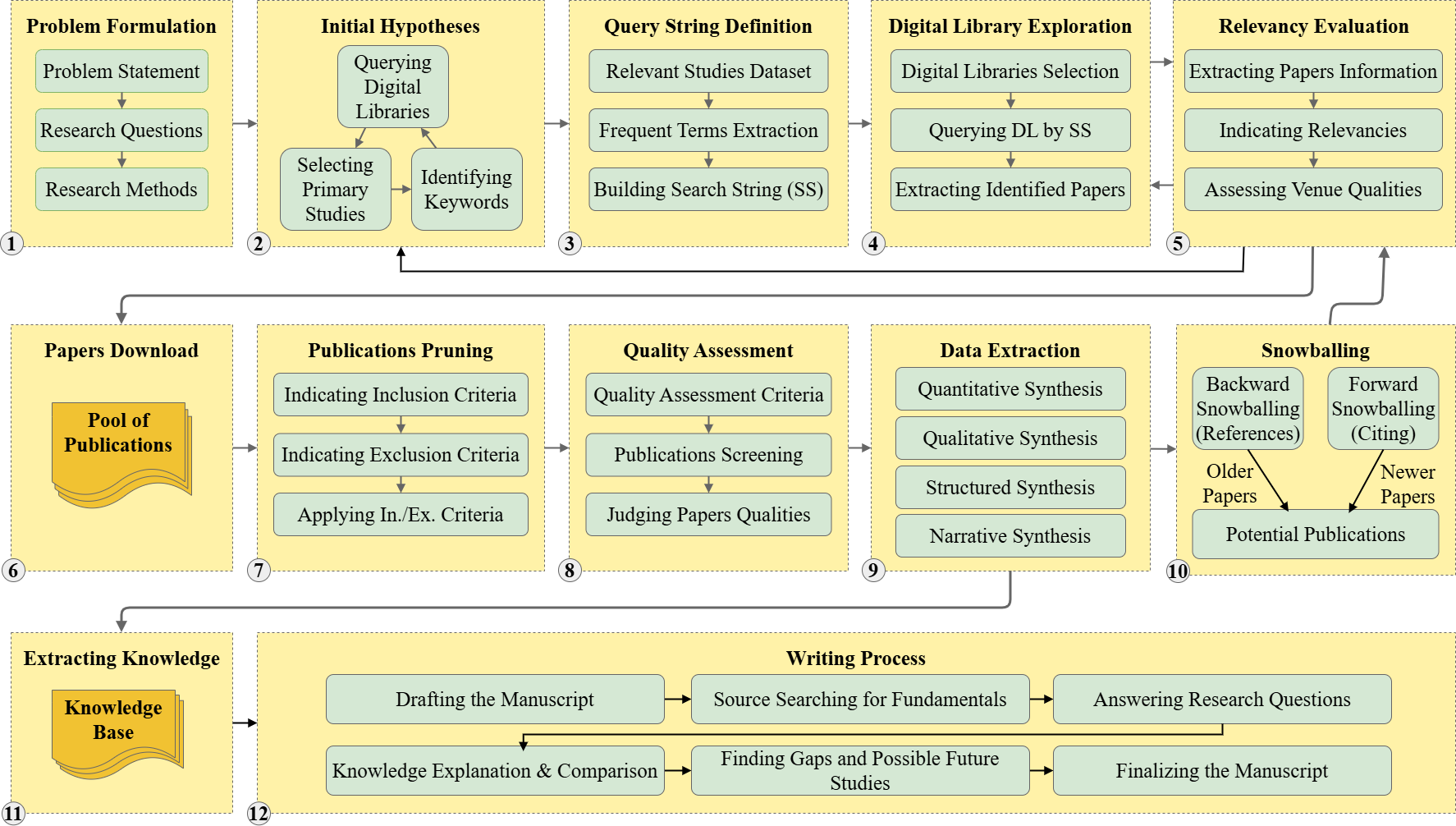}
\caption{This figure illustrates the systematic review protocol employed in this study. Developed in accordance with established methodological guidelines, the protocol consists of twelve sequential elements designed to ensure rigorous and transparent data collection and extraction from relevant literature.}\label{SLR}
\end{figure}

\begin{table}[htbp]
\centering
\small
\caption{Summary of the systematic search and selection process. The table reports the progression of study counts through the four screening phases.}
\label{tab:search_process}
\begin{tabular}{lcccc}
\hline
\textbf{Screening Phase} & \textbf{Phase 1} & \textbf{Phase 2} & \textbf{Phase 3} & \textbf{Phase 4} \\
\hline
IEEE Xplore & 650 & 45 & 15 & 12 \\
ACM Digital Library & 420 & 32 & 11 & 9 \\
Scopus & 850 & 55 & 18 & 15 \\
Web of Science & 380 & 28 & 10 & 8 \\
Springer Link & 200 & 20 & 6 & 6 \\
\hline
\textbf{Subtotal (Databases)} & \textbf{2,500} & \textbf{180} & \textbf{60} & \textbf{50} \\
\textbf{Snowballing} & \textbf{--} & \textbf{--} & \textbf{10} & \textbf{10} \\
\hline
\textbf{Total} & \textbf{2,500} & \textbf{180} & \textbf{70} & \textbf{60} \\
\hline
\end{tabular}
\end{table}
Figure \ref{SLR} presents the stepwise protocol that guided the literature search, selection, and data extraction phases of this review.

\section{Background and Theoretical Foundations}
This section establishes the foundational concepts required to understand the integration of MPC and RL. First, the core principles, mathematical formulations, and inherent limitations of MPC are reviewed. Subsequently, the fundamental elements of RL are introduced, setting the stage for analyzing how these two distinct paradigms can be synergistically combined in linear control systems.

\subsection{Model Predictive Control}
As a cornerstone of modern advanced control theory, MPC has gained widespread industrial and academic acceptance due to its explicit constraint-handling capabilities. This subsection details the mechanics of MPC, specifically focusing on linear formulations, feasibility, and stability characteristics, which serve as the baseline for evaluating hybrid MPC--RL architectures.

\subsubsection{Principles of MPC}
MPC is an advanced control strategy that utilizes numerical optimization to govern system behavior while explicitly accounting for physical, safety, and operational constraints. In this approach, the future evolution of the system is predicted over a finite look-ahead time horizon using a mathematical model, and an Optimal Control Problem (OCP) is solved in real time. At each sampling instance, only the first control action from the computed optimal sequence is applied to the plant. The horizon is then shifted forward, and the optimization process is repeated using updated system measurements--a mechanism known as the receding horizon principle. Consequently, MPC functions as a real-time, closed-loop implementation of optimal control, making it exceptionally well-suited for multi-variable systems operating near their constraint boundaries \cite{Fundamental_Rawlings2017,Fundamental_Camacho2007}.

\subsubsection{MPC Optimization Problem Formulation}
In MPC, the optimization problem is centered around minimizing a defined cost function over a finite prediction horizon $N$, relying on both the current measured state and the model-based predicted trajectory. For linear control systems, the OCP is typically cast as a Quadratic Programming (QP) problem. The standard formulation is given by:
\begin{equation} \label{eq:mpc_cost}
    \min_{U} \sum_{i=0}^{N-1} \left( \|x(k+i|k)\|_Q^2 + \|u(k+i|k)\|_R^2 \right) + \|x(k+N|k)\|_P^2
\end{equation}
subject to the linear system dynamics and constraints:
\begin{align}
    x(k+i+1|k) &= A x(k+i|k) + B u(k+i|k), \label{eq:mpc_dynamics} \\
    F x(k+i|k) &+ E u(k+i|k) \leq e, \label{eq:mpc_constraints}
\end{align}
where $x(k+i|k)$ and $u(k+i|k)$ denote the predicted state and control input at step $k+i$ given information at step $k$. The weighting matrices $Q \succeq 0$, $R \succ 0$, and the terminal penalty $P \succeq 0$ dictate the performance trade-offs. 

The inequality in \eqref{eq:mpc_constraints} represents hard limits on inputs, states, and outputs. To prevent infeasibility induced by disturbances, state constraints are frequently relaxed into soft constraints by introducing auxiliary slack variables $\epsilon(k)$. These variables allow minor constraint violations but are heavily penalized in the cost function \eqref{eq:mpc_cost} to maintain nominal operation. This explicit optimization structure is highly effective but poses significant computational demands during real-time execution, a bottleneck that learning-based approaches often seek to alleviate \cite{Fundamental_Mayne2000}.

\subsubsection{MPC Variants and Extensions}
To address diverse operational requirements, the fundamental MPC framework has been extended into several distinct variants. While linear MPC leverages linear state-space models and QP solvers for computational efficiency, nonlinear system behavior in the reviewed literature is considered only when handled through linearized predictive models within the MPC layer. Robust MPC explicitly accounts for bounded model uncertainties and disturbances--often through tube-based methods--ensuring worst-case constraint satisfaction. Alternatively, Stochastic MPC treats uncertainties probabilistically, managing risk through chance constraints. Recently, Learning-based MPC has emerged to integrate data-driven models, neural networks, or adaptive mechanisms to refine predictions and model parameters online. Collectively, these variants illustrate the adaptability of MPC, balancing computational tractability, robustness, and performance within linear or linearized predictive control formulations \cite{Fundamental_Schwenzer2021}.

\subsubsection{Constraint Handling and Feasibility}
A primary advantage of MPC over classical control methods is its systematic treatment of constraints. Actuator limits (e.g., maximum torque or voltage) are rigorously enforced as hard input constraints. Conversely, operational limits (e.g., temperature bounds) are often treated as soft state constraints to preserve the feasibility of the optimization problem in the presence of noise. To guarantee recursive feasibility--ensuring that if a solution exists at time $k$, one will also exist at time $k+1$--techniques such as robust invariant sets and tube-based strategies are deployed. These methods artificially tighten the nominal constraints during the optimization phase, guaranteeing that the true system state never violates the actual physical boundaries despite persistent disturbances \cite{Fundamental_Li2011}.

\subsubsection{Stability and Robustness}
Guaranteeing closed-loop stability is a critical requirement in control design. In MPC, stability is classically enforced through terminal ingredients: a terminal cost function (such as the matrix $P$ in \eqref{eq:mpc_cost}) and a terminal constraint set, which collectively serve as a control Lyapunov function. Robustness is further achieved by integrating uncertainty models directly into the synthesis process. However, formulating exact terminal ingredients and ensuring robustness often leads to conservative performance and increased computational complexity. Bridging this gap between rigorous theoretical guarantees (an MPC strength) and high-performance, computationally lightweight execution is the primary motivation for integrating RL into the MPC architecture \cite{Fundamental_Chen2024}.

\subsection{Reinforcement Learning}
While MPC relies on predefined mathematical models and explicit online optimization, RL approaches optimal control through a data-driven, trial-and-error paradigm. This subsection outlines the theoretical underpinnings of RL, focusing on its formulation, primary algorithmic branches, and its inherent challenges regarding safety and constraint satisfaction in physical control systems.

\subsubsection{Markov Decision Processes}
Markov Decision Processes (MDPs) provide the rigorous mathematical framework underlying sequential decision-making under uncertainty. An MDP is formally defined by the tuple $\langle \mathcal{S}, \mathcal{A}, \mathcal{P}, \mathcal{R}, \gamma \rangle$. In the context of control systems, $\mathcal{S}$ represents the continuous state space (analogous to $x_k$ in MPC), and $\mathcal{A}$ denotes the action space (analogous to the control input $u_k$). The transition dynamics are governed by $\mathcal{P}(s_{t+1}|s_t, a_t)$, representing the probability of transitioning to state $s_{t+1}$ given current state $s_t$ and action $a_t$. The agent receives a scalar feedback signal defined by the reward function $\mathcal{R}(s_t, a_t)$, while $\gamma \in [0, 1)$ is the discount factor. The objective of the RL agent is to learn an optimal policy $\pi^*(a_t|s_t)$ that maximizes the expected cumulative discounted reward, often derived by solving the Bellman optimality equations. This formulation abstracts the control problem into a framework where system dynamics are experienced rather than explicitly programmed \cite{Fundamental_Sutton1998,amirabadi2026selforganizingdualbufferadaptiveclustering}.

\subsubsection{Value-Based Methods}
Value-based RL focuses on estimating the optimal value function, which quantifies the expected long-term return from a given state or state-action pair. By accurately evaluating the action-value function, $Q(s, a)$, the agent implicitly derives its policy by greedily selecting the action that yields the highest value. Classic algorithms like Q-learning, SARSA, and their deep learning extension, Deep Q-Networks (DQN), have achieved remarkable success in environments with discrete state and action spaces. The performance of DQN are largely enabled by experience replay mechanisms, a systematic review of these mechanisms is provided in \cite{KHALILIAMIRABADI2026134403}. However, for linear control systems where the control inputs (e.g., voltages, forces) are intrinsically continuous, value-based methods suffer from the curse of dimensionality. Finding the absolute maximum of a complex $Q$-function over a continuous action space at every time step is computationally prohibitive, limiting the direct application of pure value-based methods in standard industrial control tasks \cite{Fundamental_Zai2020}.

\subsubsection{Policy-Based and Actor-Critic Methods}
To address the limitations of value-based methods in continuous domains, policy-based algorithms optimize the parameterized policy directly via gradient ascent on the expected return. Actor-Critic methods represent the state-of-the-art in this category by fusing the advantages of both policy optimization and value estimation. The ``Actor'' continuously updates the policy distribution, while the ``Critic'' evaluates the current policy to provide lower-variance gradient updates. Modern Actor-Critic algorithms such as Deep Deterministic Policy Gradient (DDPG), Twin Delayed DDPG (TD3), and Soft Actor-Critic (SAC) have become the workhorses for continuous control. They provide stable learning trajectories, robust exploration mechanisms, and are highly suitable for high-dimensional action spaces typical of multivariable control systems \cite{Fundamental_Lapan2018}.

\subsubsection{Model-Based RL}
In Model-Based Reinforcement Learning (MBRL), the agent attempts to explicitly learn the underlying transition dynamics $\mathcal{P}$ and reward function $\mathcal{R}$ to simulate and plan future trajectories. For linear control, this is conceptually akin to performing system identification on the state-space matrices $(A, B)$ concurrently with controller design. MBRL dramatically improves sample efficiency because the agent can evaluate the consequences of specific actions via simulated rollouts without requiring physical interactions. These learned predictive models can be seamlessly integrated with planning techniques or, critically, with MPC formulations. The primary vulnerability of MBRL, however, is model bias: small inaccuracies in the learned dynamics compound over long prediction horizons, leading to suboptimal or unstable control policies \cite{Fundamental_Bertsekas2019}.

\subsubsection{Safe RL and Constraints in Learning}
Standard RL algorithms are fundamentally driven by exploratory behaviors, which inherently conflict with the strict safety and operational constraints required in physical control systems. Safe RL modifies the classical MDP objective by incorporating explicit state and input constraints into the learning process. Techniques to enforce safety include reward penalty shaping, Control Barrier Functions (CBFs), shielding, and constrained policy optimization methods. While these mathematical tools provide probabilistic or deterministic safety bounds, guaranteeing recursive feasibility and strict stability during the active exploration phase remains a profound challenge. This critical gap--where RL offers superior adaptability but struggles with hard constraints, while MPC handles constraints naturally but is computationally heavy--forms the primary theoretical motivation for hybridizing MPC with RL architectures \cite{Fundamental_replaced_1, Fundamental_replaced_2, Gros2022}.

\section{Taxonomy of RL--MPC Combinations: RL Roles}
\begin{figure}[htbp]
    \centering
    \begin{tikzpicture}
        \node[anchor=south west,inner sep=0] (img) at (0,0)
        {\includegraphics[width=0.67\linewidth]{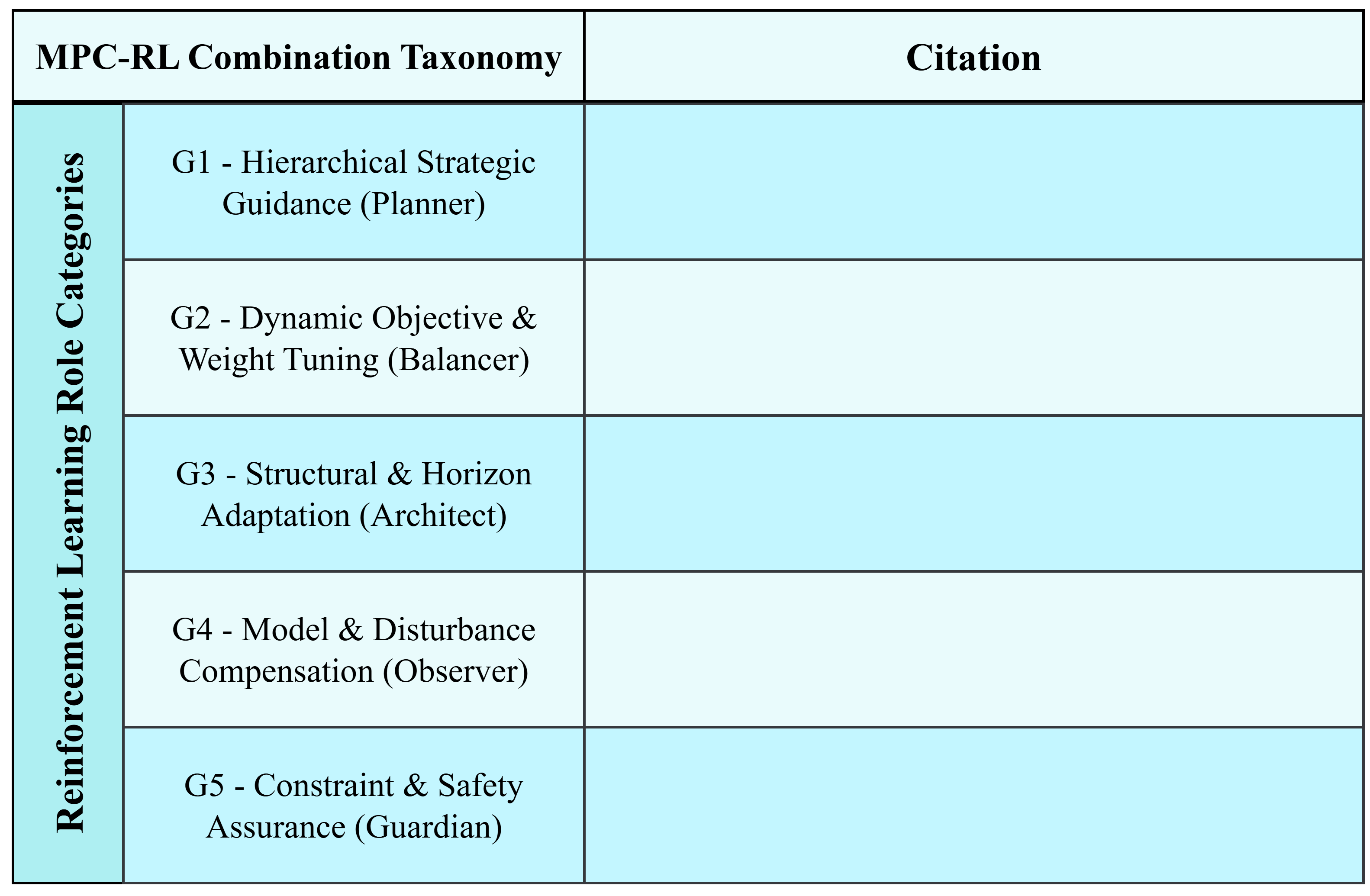}};
        \begin{scope}[x={(img.south east)},y={(img.north west)}]
            \node[align=center,font={\scriptsize}] at (0.728,0.797) {\cite{Zhao2026,Yuan2024108,Liu2024,Wu2025,Qi2024,Kim202542887,Alfonso2024141,OliveiradaSilva2025,Ren2025,Dai20224749,Xie202221163,Giannini202314830}\\\cite{Wang20247576,Capra2025,Zhang2023,Li2023,MensahAkwasi2025}};
            \node[align=center,font={\scriptsize}] at (0.728,0.623) {\cite{Jardine2019410,Jardine2021,Zhu2025,Liu2025,Gros2022,Liang2024,Peng2025,Razmi2025,Yuan2025,Qi2024,Jiang2025,Bao2024}\\\cite{Yameen2025,Fan20234763,Hu2024,Jardine20212595,NejatbakhshEsfahani2024,Zuliani20255690,Zhang20258695,Luque20241254,Wan20255353,Fan2025,Yan2025,Locher2025}\\\cite{Li202512058,Amiri2025,Chen2024964,Feng20255380,Wang2023,Usama2025,Yang2021,MensahAkwasi2025,Zhang2022,Fang2025}};
            \node[align=center,font={\scriptsize}] at (0.728,0.449) {\cite{Zhu2025,Liu2025,Hedrick2022,Peng2025,Jiang2025,Hu2024,Liu202425024,OliveiradaSilva2025,Ren2025,Zhang20258695,Cui20242217,Khalatbarisoltani202313639}\\\cite{Locher2025,Li202512058,Xie20241507,Feng20255380,Chen2024}};
            \node[align=center,font={\scriptsize}] at (0.728,0.275) {\cite{Gros2022,Zimmermann2025,Song2022,NejatbakhshEsfahani2024,Ojand202270,He202519266,Luque20241254,Wan20255353,Fu20244349,MensahAkwasi2025}};
            \node[align=center,font={\scriptsize}] at (0.728,0.101) {\cite{Gros2022,Kim202542887,NejatbakhshEsfahani2024,Zuliani20255690,Li202512058}};
        \end{scope}
    \end{tikzpicture}
    \caption{Taxonomy of Reinforcement Learning roles in MPC--RL integrations.}
    \label{Taxonomy_RL_Roles}
\end{figure}

The integration of  RL within MPC architectures is not a monolithic paradigm; rather, it manifests across a spectrum of configurations depending on which component of the optimal control problem the learning agent targets. In modern frameworks applied to linear control systems, RL functions not as a replacement for MPC, but as a specialized augmentation module. By systematically analyzing the literature, we categorize the complementary roles of RL into five distinct architectural archetypes, as comprehensively illustrated in the taxonomy diagram in Figure \ref{Taxonomy_RL_Roles}:

\begin{itemize}
    \item \textbf{Hierarchical Strategic Guidance (Planner):}  
    In this cascading architecture, RL operates in the outer loop as a high-level strategic supervisor, operating on a slower timescale than the local controller. Its primary functions include generating optimal reference trajectories, selecting discrete operating modes, and arbitrating between competing long-term objectives. Crucially, in this configuration, RL is strictly confined to generating these supervisory commands and does not intervene in the internal optimization mechanics of the MPC. Because the lower-level MPC layer is left entirely intact, it remains solely responsible for high-frequency reference tracking and disturbance rejection. This structural decoupling guarantees that the original cost function, physical constraints, and rigorous stability proofs (e.g., recursive feasibility) of the underlying linear MPC formulation are preserved.

    \item \textbf{Dynamic Objective and Weight Tuning (Balancer):}  
    Manual tuning of the MPC cost function is notoriously unintuitive, especially in multi-objective scenarios. In the Balancer role, the RL agent dynamically adapts the optimization landscape by continuously tuning the state and input weighting matrices ($Q$ and $R$), the terminal penalty matrix ($P$), or by learning complex terminal cost functions. By observing system performance over time, the RL agent actively manages real-time trade-offs--such as aggressive reference tracking versus energy conservation, or transient performance versus actuator wear. This dynamic mapping of operating conditions to optimal weights expands the region of attraction and enhances the closed-loop adaptability without altering the linear prediction model.

    \item \textbf{Structural and Horizon Adaptation (Architect):}  
    Real-time implementation of MPC is fundamentally bottlenecked by the computational burden of solving mathematical programs at every sampling instant. The Architect role leverages RL to dynamically modify the structural complexity of the MPC problem to guarantee real-time feasibility. The RL agent observes the computational limits and system state to adaptively shrink or expand the prediction horizon ($N$) and control horizon ($N_c$). Furthermore, RL can be utilized to generate highly accurate initial guesses for the optimization variables (warm-starting), which significantly accelerates the convergence rate of active-set or interior-point solvers, thereby balancing theoretical optimality with strict computational efficiency.

\item \textbf{Model and Disturbance Compensation (Observer):}  
Despite the robustness of MPC, significant plant-model mismatches in linear approximations inevitably degrade control performance. Acting as a data-driven estimator, the RL agent addresses the ``reality gap'' by learning residual model mismatch around nominal linear or linearized predictive models, estimating additive transient disturbances ($d_k$), or capturing parametric drifts in the nominal state-space matrices ($A$ and $B$). Such model augmentation can be executed in two primary paradigms: \textit{online learning}, enabling continuous real-time adaptation to changing environments and exogenous disturbances; or \textit{offline learning}, where historical operation data is utilized to refine the nominal model prior to deployment. In both cases, these learned residual components augment the linear or linearized MPC model, improving its resilience against environmental uncertainty.

 \item \textbf{Constraint and Safety Assurance (Guardian):}  
    While standard RL struggles with strict bounds due to its exploratory nature, MPC excels at explicit constraint handling. However, deterministic constraints in MPC can lead to infeasibility under severe uncertainties. In the Guardian role, RL enhances system safety by adaptively managing the constraint space. It achieves this through dynamic constraint tightening (similar to tube-based MPC concepts), tuning the penalty weights of slack variables for soft constraints, or acting as a safety shield/filter that modifies aggressive baseline actions. Notably, while the RL agent acts as a safety supervisor by shaping the feasible region, the final control decision remains entirely under the authority of the MPC optimization layer, ensuring that critical physical limits are not breached during operation.
\end{itemize}

\begin{figure}[htbp]
    \centering
    \includegraphics[width=0.5\linewidth]{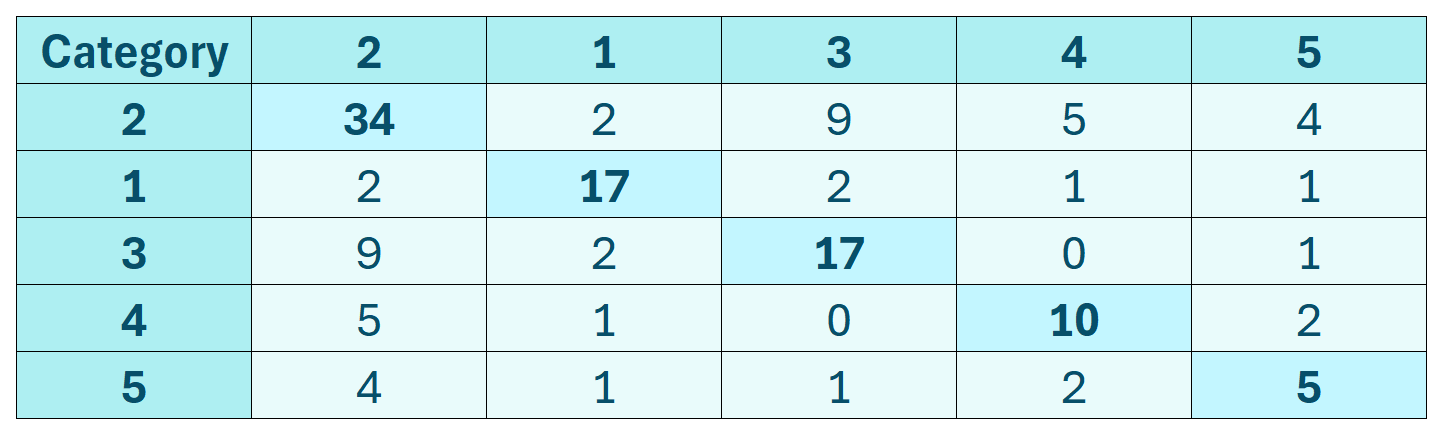}
    \caption{The co-occurrence between categories based on RL roles in MPC--RL integrations.}
    \label{RL_ROLE_Correlations}
\end{figure}

\begin{figure}[htbp]
    \centering
    \includegraphics[width=0.65\linewidth]{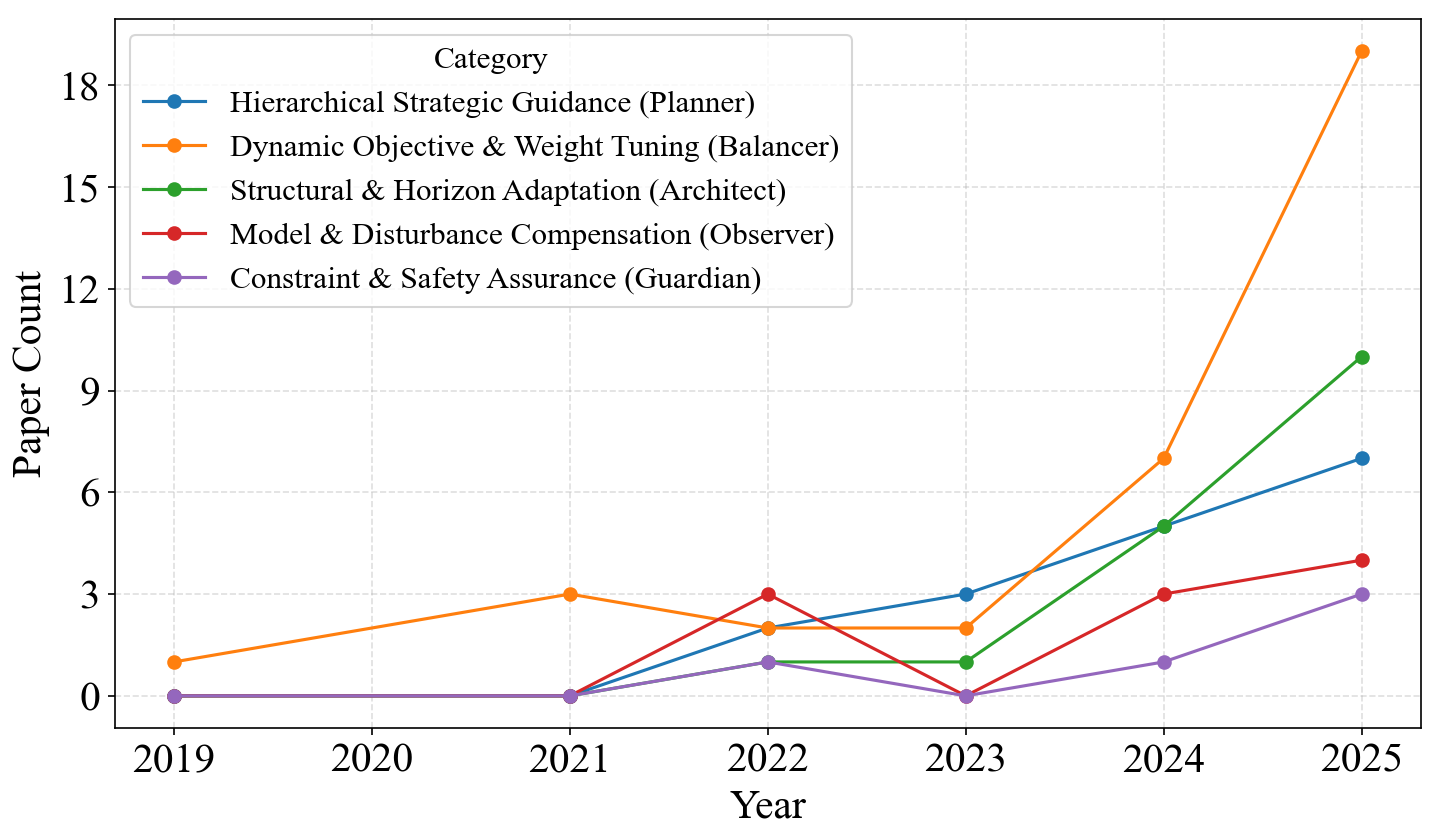}
    \caption{Temporal distribution of RL roles applied in MPC--RL integrations.}
    \label{RL_ROLE_Year}
\end{figure}

To provide deeper insight into the interplay of these configurations, Figure \ref{RL_ROLE_Correlations} visualizes the co-occurrence and structural overlaps among the defined categories. The literature reveals a growing trend where modern frameworks do not strictly isolate these functions; instead, they often employ hybrid roles (e.g., simultaneously acting as an Observer for model compensation and a Balancer for weight tuning) to maximize closed-loop performance. Furthermore, the temporal distribution of these applied roles across the reviewed literature is presented in Figure \ref{RL_ROLE_Year}. This chronological analysis highlights a recent and significant paradigm shift: while early research predominantly focused on high-level Planner roles, recent years have witnessed a massive surge in Observer and Balancer applications, driven by advances in deep RL and the critical need for real-time adaptability in complex linear systems.

\section{Taxonomy of RL--MPC Combinations: RL Algorithm Types}
\begin{figure}[htbp]
    \centering
    \begin{tikzpicture}
        \node[anchor=south west,inner sep=0] (img) at (0,0)
        {\includegraphics[width=0.5\linewidth]{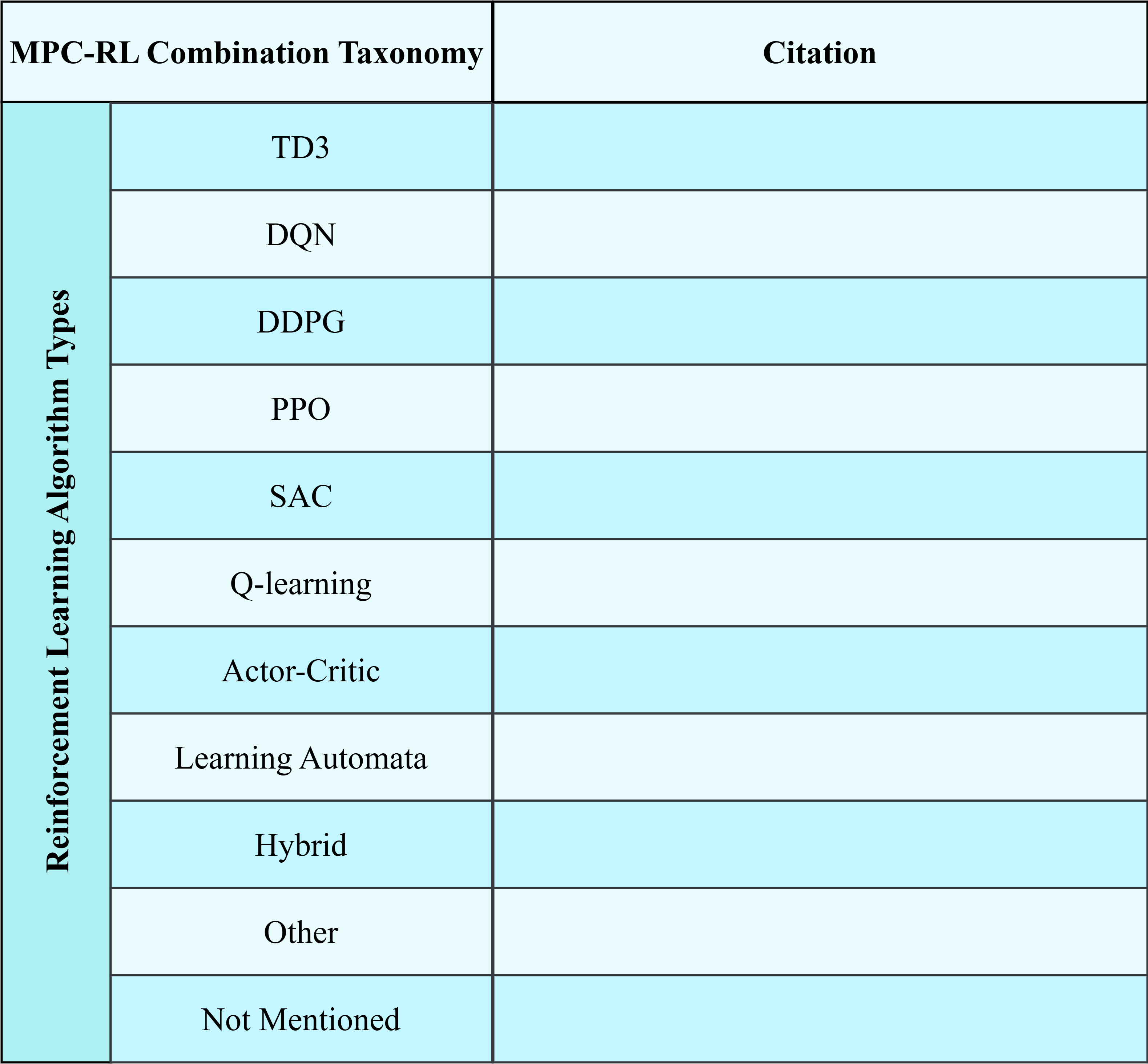}};
        \begin{scope}[x={(img.south east)},y={(img.north west)}]
            \node[align=center,font={\scriptsize}] at (0.712,0.855) {\cite{Cui20242217,Zhu2025,Liu2025,Liu2024,Peng2025,Jiang2025,Hu2024,Wang20247576,Usama2025}};
            \node[align=center,font={\scriptsize}] at (0.712,0.768) {\cite{Liang2024,Yuan2024108,Alfonso2024141,Liu202425024,OliveiradaSilva2025,Fan2025,Fu20244349,Zhang2022,Fang2025}};
            \node[align=center,font={\scriptsize}] at (0.712,0.690) {\cite{Zhao2026,Wu2025,Fan20234763,Xie20241507,Feng20255380,Li2023}};
            \node[align=center,font={\scriptsize}] at (0.712,0.612) {\cite{Chen2024964,Yan2025,Capra2025,Zhang2023,Chen2024}};
            \node[align=center,font={\scriptsize}] at (0.712,0.533) {\cite{Kim202542887,He202519266,Li202512058,Wang2023}};
            \node[align=center,font={\scriptsize}] at (0.712,0.455) {\cite{Song2022,Yuan2025,Ojand202270,Giannini202314830,Amiri2025}};
            \node[align=center,font={\scriptsize}] at (0.712,0.377) {\cite{Zimmermann2025,Dai20224749}};
            \node[align=center,font={\scriptsize}] at (0.712,0.299) {\cite{Jardine20212595,Jardine2021,Jardine2019410}};
            \node[align=center,font={\scriptsize}] at (0.712,0.211) {\cite{Wan20255353,Zhang20258695,MensahAkwasi2025}};
            \node[align=center,font={\scriptsize}] at (0.712,0.133) {\cite{Qi2024,NejatbakhshEsfahani2024,Yameen2025,Khalatbarisoltani202313639,Zuliani20255690,Hedrick2022}};
            \node[align=center,font={\scriptsize}] at (0.712,0.055) {\cite{Locher2025,Yang2021,Gros2022,Bao2024,Ren2025,Luque20241254,Xie202221163,Razmi2025}};
        \end{scope}
    \end{tikzpicture}
    \caption{Taxonomy of Reinforcement Learning algorithms implemented in MPC--RL frameworks.}
    \label{Taxonomy_RL_Types}
\end{figure}

To provide a coherent analysis of the diverse RL methods employed in MPC--RL frameworks, the learning algorithms identified in the surveyed literature are systematically grouped into standardized categories. As illustrated in the detailed taxonomy diagram in Figure \ref{Taxonomy_RL_Types}, this categorization is based on the underlying learning paradigm, policy representation, and optimization mechanism. By mapping these algorithms, we reduce terminology ambiguity and provide clear insights into why specific RL architectures are selected for particular control objectives (e.g., continuous parameter tuning versus discrete mode switching). 

The prominent RL algorithms utilized in conjunction with MPC are categorized as follows:

\begin{itemize}
    \item \textbf{TD3 (Twin Delayed Deep Deterministic Policy Gradient):}  
    An advanced actor--critic algorithm designed for continuous action spaces. In MPC applications, TD3 is highly favored for tuning continuous variables (e.g., weights or prediction horizons) because it effectively mitigates the overestimation bias inherent in standard Q-learning, providing highly stable and precise parameter adjustments.
	
    \item \textbf{DQN (Deep Q-Network):}  
    A value-based method inherently designed for discrete action spaces. In the context of MPC, DQN and its variants (such as Double DQN, REDQ, and VDN) are predominantly utilized in high-level \textit{Planner} roles, where the RL agent makes discrete decisions, such as selecting between predefined operating modes, switching local MPC controllers, or deciding binary fault-recovery actions.
	
    \item \textbf{DDPG (Deep Deterministic Policy Gradient):}  
    A pioneering off-policy actor--critic algorithm for continuous control. While historically significant and widely implemented in early MPC--RL integrations for continuous parameter mapping, it is increasingly being superseded by more stable variants like TD3 and SAC due to its sensitivity to hyperparameter tuning.
	
    \item \textbf{PPO (Proximal Policy Optimization):}  
    An on-policy algorithm celebrated for its stable and conservative policy updates via trust-region clipping. In control systems where sudden, aggressive changes to MPC parameters could destabilize the plant, PPO (along with variants like OSPPO) offers a safer learning trajectory, albeit at the cost of lower sample efficiency compared to off-policy methods.
	
    \item \textbf{SAC (Soft Actor--Critic):}  
    An off-policy algorithm founded on the maximum entropy RL framework. SAC inherently balances exploration and exploitation, making it exceptionally robust against plant-model mismatches. It is increasingly utilized in \textit{Observer} and \textit{Balancer} roles where the MPC operates in highly stochastic or uncertain environments.
	
    \item \textbf{Q-learning:}  
    Standard tabular or function-approximated Q-learning methods (including variants like Federated QAvg). Distinct from deep Q-networks, these are typically employed in simpler, lower-dimensional state spaces or when computational resources are strictly limited.
	
    \item \textbf{Actor--Critic (General):}  
    Broad applications of the actor--critic architecture that do not explicitly align with the highly parameterized variants (like DDPG or TD3), often tailored specifically to the unique dynamics of the target linear system.
	
    \item \textbf{Learning Automata:}  
    Probability-updating algorithms, including Fixed-Structure Learning Automata (FALA) and generic LA. These are typically applied in stochastic environments for simple, adaptive decision-making alongside the MPC.
	
    \item \textbf{Hybrid Approaches:}  
    Frameworks that simultaneously leverage multiple distinct RL algorithms (e.g., combining DDPG and TD3, or Q-learning with an actor--critic structure) to handle mixed discrete-continuous action spaces or multi-agent hierarchical control tasks.
	
    \item \textbf{Other Architectures:}  
    Less frequently implemented approaches in the MPC--RL domain, including SARSA, Offline RL, Deterministic Policy Gradient (DPG), Relative Entropy Policy Search (REPS), and Adaptive Dynamic Programming (ADP).
	
    \item \textbf{Not Mentioned:}  
    Studies that conceptually utilize RL as a black-box optimizer or adaptive mechanism without explicitly specifying the underlying algorithmic architecture.
\end{itemize}

\begin{figure}[htbp]
    \centering
    \includegraphics[width=0.61\linewidth]{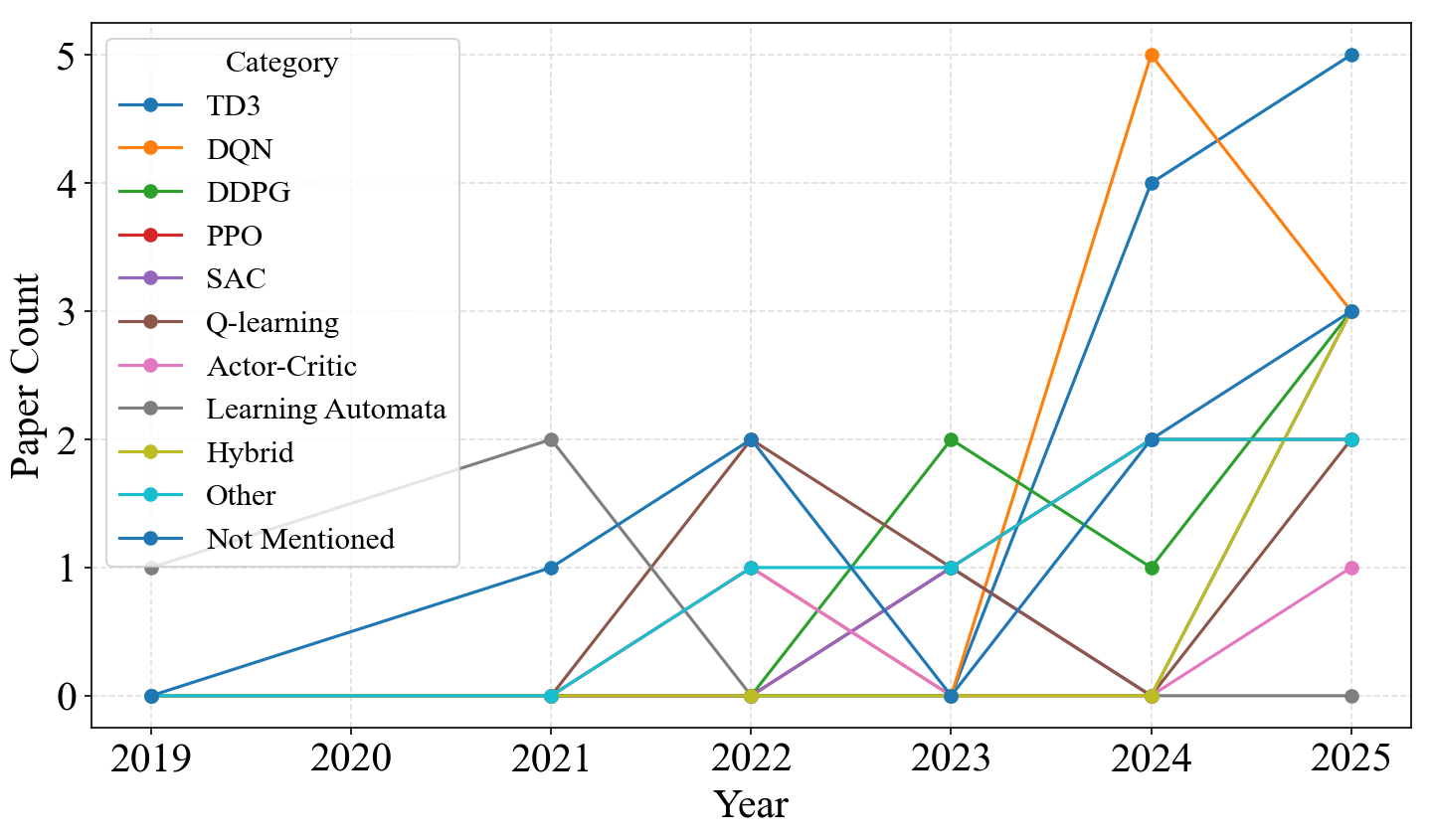}
    \caption{Temporal evolution of RL algorithm types adopted in MPC--RL literature.}
    \label{RL_Types_Year}
\end{figure}

To understand the evolution of these algorithmic choices, Figure \ref{RL_Types_Year} presents the temporal distribution of RL types applied within MPC--RL frameworks. A critical analysis of this timeline reveals a paradigm shift. Early research heavily relied on value-based and discrete methods (such as standard Q-learning). However, as the complexity of linear control applications increased, there has been an exponential surge in the adoption of advanced, continuous deep RL algorithms--most notably TD3, SAC, and PPO. This trend underscores a growing consensus in the field: modern MPC augmentation requires the high-dimensional, continuous mapping capabilities that only state-of-the-art actor--critic architectures can provide.


\section{Taxonomy of RL--MPC Combinations: Cost Function Types}
\label{sec:taxonomy_cost_function}

The cost function, or objective function, lies at the heart of any MPC formulation. It mathematically encapsulates the control objectives, performance criteria, and operational constraints that the controller must optimize. In the context of RL--MPC, the design of this function is a critical nexus where classical control theory objectives meet data-driven policies learned by RL. This section provides a comprehensive, multi-dimensional taxonomy of cost functions used in the RL--MPC literature, deconstructing their formulation based on control objectives, temporal structure, penalized variables, and other key characteristics. This structured analysis illuminates the underlying design trade-offs and research trends in the field.

\subsection{Classification by Control Objective}
The highest-level classification of a cost function pertains to its fundamental control goal. We identify three primary categories:
\begin{itemize}
\item \textbf{Regulation (Setpoint Stabilization):} This is the most traditional control objective, representing 8.3\% (5 out of 60) of the reviewed literature. It aims to drive the system states to a desired equilibrium or setpoint (often the origin, $x_{sp}=0, u_{sp}=0$) and maintain them there in the face of disturbances. The cost function penalizes deviations of the state and/or control input from this setpoint. Its primary goal is ensuring stability and disturbance rejection.

\item \textbf{Tracking (Trajectory Following):} In this formulation, which constitutes the vast majority of the surveyed articles at 83.3\% (50 out of 60), the objective is to make the system's output or state follow a predefined, time-varying reference trajectory, $r(t)$. The cost function is structured to penalize the tracking error, i.e., the difference between the system's trajectory and the reference trajectory. This is common in applications like robotics and process control where a specific path or profile must be followed.
    
 \item \textbf{Economic (or General Performance):} Rather than tracking a specific setpoint or trajectory, economic MPC (EMPC) (accounting for the remaining 8.3\%, or 5 out of 60 articles) aims to optimize a direct measure of system performance or economic benefit over time. This can include minimizing energy consumption, maximizing production rate, or reducing operational costs. The cost function in this case is not necessarily positive definite with respect to a specific equilibrium, posing unique challenges and opportunities for ensuring stability while optimizing for broader performance goals.
\end{itemize}

\subsection{Classification by Temporal Structure}
Beyond the primary goal, the temporal structure of the cost function dictates how performance is evaluated over the prediction horizon. It is typically composed of two components:

\begin{itemize}
\item \textbf{Stage Cost ($L$):} This term, denoted $L(x_k, u_k)$, is evaluated at each step $k$ within the prediction horizon $N$. It quantifies the instantaneous cost associated with the system's state and control input at that time step. A significant majority of the reviewed literature, 61.7\% (37 out of 60 articles), employs a ``stage-only'' formulation, where the total cost is exclusively a summation of these instantaneous costs.
    
\item \textbf{Terminal Cost ($F$):} This term, denoted $F(x_N)$, is evaluated only at the end of the prediction horizon. It is incorporated in the remaining 38.3\% (23 out of 60) of the surveyed works. In traditional MPC, the terminal cost is a critical design element used to guarantee the stability of the closed-loop system. By appropriately choosing $F(\cdot)$ (often as the solution to the Lyapunov equation), one can ensure that the controller is recursively feasible and the system state remains bounded. In many RL--MPC schemes, this term might be learned or approximated by a value function from the RL agent.
\end{itemize}

The total cost is the summation of the stage costs over the horizon, plus the terminal cost: $J = \sum_{k=0}^{N-1} L(x_k, u_k) + F(x_N)$.

\subsection{Classification by Penalized Variables}
The choice of variables included in the stage and terminal costs directly reflects what aspects of system behavior are being controlled. Common choices include:
\begin{itemize}
    \item \textbf{State-based} State-based costs penalize the full system state vector. This formulation, utilized in 41.7\% (25 out of 60) of the reviewed works, provides strong stability properties and a clear physical interpretation but requires accurate state measurement or estimation.
    \item \textbf{Output-based} Output-based costs penalize deviations in system outputs rather than states. This approach, found in 55.0\% (33 out of 60) of the literature, aligns naturally with performance variables and industrial tuning practices but offers indirect state regulation.
    \item \textbf{Input-Magnitude-based} Input costs penalize the magnitude or energy of control inputs. As the most frequently included individual term, it appears in 63.3\% (38 out of 60) of the articles. These terms prevent excessive actuation and are essential for systems with actuator limitations.
    \item \textbf{Input-Rate-of-Change-based ($\Delta u$)} Input increment costs penalize changes in control inputs between successive time steps. This term, present in 36.7\% (22 out of 60) of cases, promotes smooth control actions, reduces actuator wear, and is widely used in industrial MPC.
    \item \textbf{Mixed or Combined Formulations} Mixed cost functions combine penalties on states, outputs, inputs, and input increments. This formulation provides flexibility in shaping system behavior. Confirming its dominance, our analysis shows that 90.0\% (54 out of 60) of the reviewed articles employ a mixed structure, penalizing at least two of the aforementioned variable classes.
 
\end{itemize}
Penalizing input rates ($\Delta u = u_k - u_{k-1}$) is particularly important in practical applications to prevent excessive wear on actuators and to ensure smoother control actions.

\textbf{Note:}
A specific category is dedicated to Input Reference Tracking, which involves penalizing the deviation from a desired input reference, $\|u - u_{ref}\|^2$. This highly specific formulation is quite rare in the reviewed literature, observed in only 1.7\% (1 out of 60) of the works.

\subsection{Classification by Weighting Structure}
The relative importance of different objectives and time steps is managed through weighting matrices.
\begin{itemize}
    \item \textbf{Uniform Weighting:} Constant weighting matrices ($Q, R$) are used for all steps in the horizon. This is the most prevalent approach, accounting for a significant majority of 70.0\% (42 out of 60) of the surveyed articles.
    \item \textbf{Time-Varying Weighting:} The weights ($Q_k, R_k$) change over the horizon, allowing for different priorities at different times. This dynamic weighting strategy is utilized in 26.7\% (16 out of 60) of the reviewed literature.
    \item \textbf{Discounted Cost:} A discount factor $\gamma \in (0, 1]$ is applied, i.e., $\sum_{k=0}^{N-1} \gamma^k L(x_k, u_k)$. This is standard in RL and prioritizes short-term performance over long-term, which can be beneficial for infinite-horizon problems. Interestingly, despite its foundational role in general RL, explicitly discounted formulations appear in only 3.3\% (2 out of 60) of the analyzed RL--MPC works.
\end{itemize}

\subsection{Classification by Constraint Handling}
MPC's ability to handle constraints is a primary advantage. Constraints can be enforced in two ways, and this review shows a clear preference in how they are applied:
\begin{itemize}
    \item \textbf{Hard Constraints:} Formulated as strict inequalities in the optimization problem. Violations are not permitted. This strict enforcement is the dominant approach in the literature, utilized by a significant 73.3\% (44 out of 60) of the surveyed works.
    \item \textbf{Soft Constraints:} Flexibility is introduced by allowing constraints to be violated at a certain cost. This is achieved by introducing non-negative ``slack'' variables ($\varepsilon$) into the constraints (e.g., $h(x) \leq \varepsilon$) and adding a penalty term for these variables to the cost function (e.g., $\rho\|\varepsilon\|^2$). This enhances robustness by preventing infeasibility when constraints cannot be perfectly met. This slack-penalized formulation is adopted by the remaining 26.7\% (16 out of 60) of the reviewed articles, typically in scenarios where preventing solver failure is more critical than strict boundary adherence.
\end{itemize}

\subsection{Classification by Cost Type}
The mathematical structure of the cost function is paramount as it directly determines the complexity of the resulting optimization problem. This analysis categorizes the reviewed literature into three main structural classes:
\begin{itemize}
    \item \textbf{Quadratic:} 
    The dominant pattern is the standard MPC/LQR-style quadratic tracking/regulation functional composed of weighted squared norms of (i) state/output tracking error and (ii) control effort, often augmented with (iii) move-suppression terms (e.g., $\|\Delta u\|^2$) and (iv) terminal penalties.
A generic representative form is
\begin{equation}
	\label{eq:quad_generic}
	J \;=\; \sum_{k=0}^{N-1}
	\Big(
	\|y_k-r_k\|_{Q}^{2}
	+\|x_k-x_k^{\mathrm{ref}}\|_{Q_x}^{2}
	+\|u_k\|_{R}^{2}
	+\|\Delta u_k\|_{S}^{2}
	\Big)
	\;+\;
	\|x_N-x_N^{\mathrm{ref}}\|_{P}^{2},
\end{equation}
where $Q,Q_x,R,S,P\succeq 0$ encode relative priorities. This class is preferred because it yields convex QP structure for linear models and retains tractability under common convexifications for nonlinear MPC. 
%
%
     Confirming its widespread preference, an overwhelming 90.0\% (54 out of 60) of the surveyed works employ a quadratic or quadratic-dominant structure.

  \item \textbf{Linear/Economic:}
A smaller subset uses explicitly economic/utility-oriented formulations, in which the objective is expressed as (weighted) sums of key performance indicators and may include explicit maximization:
\begin{equation}
	\label{eq:econ_generic}
	J \;=\; \max \;\sum_{k=0}^{N-1} w^\top \psi(x_k,u_k),
	\qquad\text{or equivalently}\qquad
	\min \; -\sum_{k=0}^{N-1} w^\top \psi(x_k,u_k),
\end{equation}
where $\psi(\cdot)$ collects revenue/cost terms (energy, throughput, comfort, etc.). These formulations are common in economic MPC and application-driven MPC where the natural metric is profit/cost rather than squared tracking error. This class represents a small fraction of the literature, observed in just 5.0\% (3 out of 60) of the articles.
    \item \textbf{Mixed (Non-Quadratic):}
This category encompasses formulations resulting in a general, non-convex Nonlinear Program (NLP). While highly expressive and capable of handling complex objectives, NLPs are computationally demanding and may converge to local minima. A prominent formulation within this class involves mixed objectives that combine standard quadratic terms ($J_{\mathrm{quad}}$) with non-quadratic elements such as hinge/rectifier penalties, max operators, or other non-smooth components ($\Phi$):
\begin{equation}
	\label{eq:mixed_generic}
	J \;=\; J_{\mathrm{quad}}(x,u)
	\;+\; \lambda \sum_{k=0}^{N-1} \max\{0,\ g(x_k,u_k)\}
	\;+\; \eta \,\Phi(x,u),
\end{equation}
where the $\max\{0,\cdot\}$ term captures one-sided penalties (e.g., safety margins). Such objectives typically arise when the performance metric is inherently piecewise (comfort/safety thresholds, saturation-aware penalties) or when robustness is embedded directly into the objective. Due to their complexity, these advanced formulations are utilized in only 5.0\% (3 out of 60) of the reviewed papers.
\end{itemize}

\subsection{Safety in Cost Function}
Integrating safety requirements directly into the objective function is a critical design choice in control and learning architectures. Rather than enforcing safety exclusively through hard constraints, some formulations embed safety-awareness into the cost function via penalty shaping, soft constraints, or slack variables. Based on the reviewed literature, the approaches can be divided into two distinct categories regarding the inclusion of safety terms in the cost function:
\begin{itemize}   
 \item \textbf{Cost with Safety Considerations (18.3\%, 11 out of 60):} A minority of the analyzed works explicitly embed safety considerations into the objective function. These formulations typically utilize penalty weights on slack variables or barrier-like penalty terms to create a ``soft'' boundary. This approach actively penalizes unsafe states or aggressive control actions, guiding the system away from hazardous regions while maintaining the feasibility of the optimization problem.
    \item \textbf{ Cost without Safety Terms (81.7\%, 49 out of 60):} The vast majority of the reviewed literature does not incorporate specific safety penalties directly into the cost function. Instead, these works rely on standard tracking, regulation, or economic objectives. In such architectures, safety is typically managed independently, either by enforcing strict hard constraints within the MPC formulation or by employing external safety filters and shielding mechanisms outside the primary objective.\end{itemize}

\begin{figure}[!h]
	\centering
	\begin{tikzpicture}
		\node[anchor=south west,inner sep=0] (img) at (0,0)
		{\includegraphics[width=0.98\linewidth]{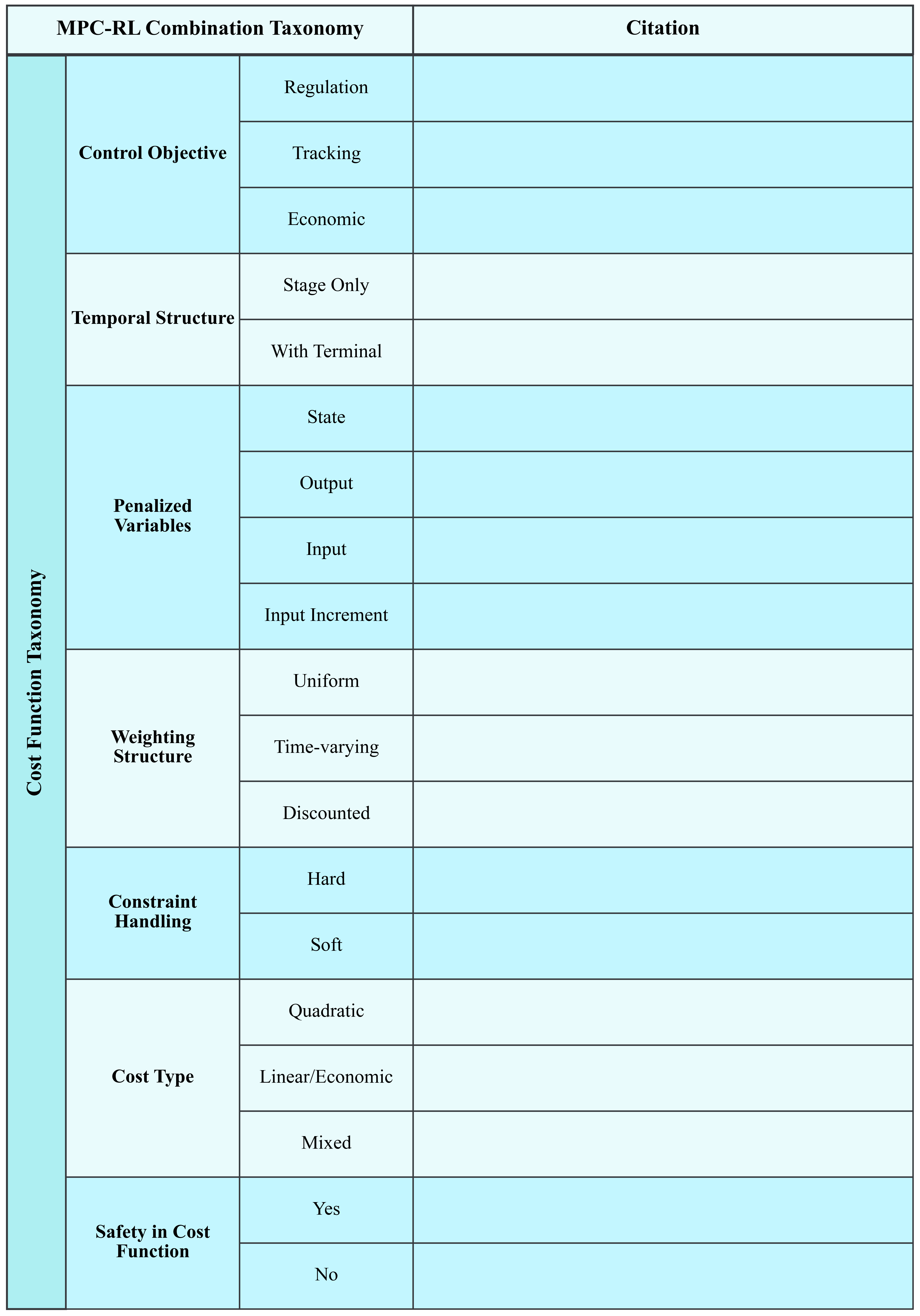}};
		\begin{scope}[x={(img.south east)},y={(img.north west)}]
			\node[align=center,font={\scriptsize}] at (0.721,0.933) {\cite{Song2022,Qi2024,Fan20234763,Zuliani20255690,Amiri2025}};
			\node[align=center,font={\scriptsize}] at (0.721,0.883) {\cite{Jardine2019410,Jardine2021,Zhu2025,Gros2022,Hedrick2022,Liang2024,Zhao2026,Yuan2024108,Liu2024,Peng2025,Wu2025,Zimmermann2025,Razmi2025,Yuan2025,Jiang2025,Bao2024,Yameen2025,Hu2024}\\\cite{Kim202542887,Alfonso2024141,Liu202425024,Jardine20212595,NejatbakhshEsfahani2024,Ren2025,Zhang20258695,He202519266,Cui20242217,Luque20241254,Wan20255353,Dai20224749,Fan2025,Xie202221163,Giannini202314830,Yan2025,Locher2025,Wang20247576}\\\cite{Li202512058,Xie20241507,Chen2024964,Feng20255380,Wang2023,Capra2025,Zhang2023,Usama2025,Li2023,Yang2021,Chen2024,MensahAkwasi2025,Zhang2022,Fang2025}};
			\node[align=center,font={\scriptsize}] at (0.721,0.832) {\cite{Liu2025,OliveiradaSilva2025,Ojand202270,Khalatbarisoltani202313639,Fu20244349}};
			\node[align=center,font={\scriptsize}] at (0.721,0.782) {\cite{Zhu2025,Liu2025,Liang2024,Zhao2026,Yuan2024108,Liu2024,Wu2025,Razmi2025,Yuan2025,Bao2024,Yameen2025,Fan20234763,OliveiradaSilva2025,Ojand202270,Zhang20258695,He202519266,Cui20242217,Wan20255353}\\\cite{Dai20224749,Xie202221163,Giannini202314830,Fu20244349,Wang20247576,Li202512058,Amiri2025,Xie20241507,Chen2024964,Feng20255380,Wang2023,Capra2025,Zhang2023,Usama2025,Li2023,Yang2021,Chen2024,Zhang2022}\\\cite{Fang2025}};
			\node[align=center,font={\scriptsize}] at (0.721,0.732) {\cite{Jardine2019410,Jardine2021,Gros2022,Hedrick2022,Peng2025,Zimmermann2025,Song2022,Qi2024,Jiang2025,Hu2024,Kim202542887,Alfonso2024141,Liu202425024,Jardine20212595,NejatbakhshEsfahani2024,Zuliani20255690,Ren2025,Luque20241254}\\\cite{Fan2025,Khalatbarisoltani202313639,Yan2025,Locher2025,MensahAkwasi2025}};
			\node[align=center,font={\scriptsize}] at (0.721,0.682) {\cite{Gros2022,Zhao2026,Liu2024,Song2022,Yuan2025,Qi2024,Fan20234763,Hu2024,Kim202542887,Alfonso2024141,Liu202425024,NejatbakhshEsfahani2024,Zuliani20255690,Ren2025,Wan20255353,Khalatbarisoltani202313639,Giannini202314830,Yan2025}\\\cite{Li202512058,Amiri2025,Capra2025,Zhang2023,Li2023,Yang2021,Fang2025}};
			\node[align=center,font={\scriptsize}] at (0.721,0.632) {\cite{Jardine2019410,Jardine2021,Zhu2025,Liu2025,Hedrick2022,Liang2024,Yuan2024108,Peng2025,Wu2025,Zimmermann2025,Razmi2025,Jiang2025,Bao2024,Yameen2025,Jardine20212595,Zhang20258695,He202519266,Cui20242217}\\\cite{Luque20241254,Wan20255353,Dai20224749,Fan2025,Xie202221163,Locher2025,Wang20247576,Xie20241507,Chen2024964,Feng20255380,Wang2023,Usama2025,Chen2024,MensahAkwasi2025,Zhang2022}};
			\node[align=center,font={\scriptsize}] at (0.721,0.582) {\cite{Jardine2019410,Jardine2021,Liu2025,Gros2022,Zhao2026,Yuan2024108,Liu2024,Wu2025,Song2022,Yuan2025,Qi2024,Yameen2025,Fan20234763,Hu2024,Kim202542887,Alfonso2024141,Liu202425024,Jardine20212595}\\\cite{NejatbakhshEsfahani2024,OliveiradaSilva2025,Zuliani20255690,Ren2025,Ojand202270,Wan20255353,Dai20224749,Khalatbarisoltani202313639,Giannini202314830,Yan2025,Fu20244349,Locher2025,Li202512058,Amiri2025,Wang2023,Capra2025,Zhang2023,Usama2025}\\\cite{Li2023,Zhang2022}};
			\node[align=center,font={\scriptsize}] at (0.721,0.532) {\cite{Zhu2025,Hedrick2022,Liang2024,Yuan2024108,Peng2025,Zimmermann2025,Yuan2025,Jiang2025,Bao2024,Zhang20258695,He202519266,Luque20241254,Fan2025,Xie202221163,Wang20247576,Xie20241507,Chen2024964,Feng20255380}\\\cite{Wang2023,Yang2021,Chen2024,Fang2025}};
			\node[align=center,font={\scriptsize}] at (0.721,0.482) {\cite{Jardine2019410,Jardine2021,Gros2022,Hedrick2022,Zhao2026,Yuan2024108,Liu2024,Wu2025,Zimmermann2025,Song2022,Qi2024,Jiang2025,Yameen2025,Fan20234763,Hu2024,Kim202542887,Alfonso2024141,Jardine20212595}\\\cite{Zuliani20255690,Ren2025,He202519266,Cui20242217,Wan20255353,Dai20224749,Fan2025,Xie202221163,Khalatbarisoltani202313639,Giannini202314830,Fu20244349,Wang20247576,Li202512058,Xie20241507,Chen2024964,Feng20255380,Wang2023,Capra2025}\\\cite{Zhang2023,Li2023,Yang2021,Chen2024,MensahAkwasi2025,Fang2025}};
			\node[align=center,font={\scriptsize}] at (0.721,0.431) {\cite{Zhu2025,Liu2025,Liang2024,Peng2025,Razmi2025,Yuan2025,Bao2024,OliveiradaSilva2025,Ojand202270,Zhang20258695,Luque20241254,Yan2025,Locher2025,Amiri2025,Usama2025,Zhang2022}};
			\node[align=center,font={\scriptsize}] at (0.721,0.381) {\cite{Liu202425024,NejatbakhshEsfahani2024}};
			\node[align=center,font={\scriptsize}] at (0.721,0.331) {\cite{Jardine2019410,Jardine2021,Zhu2025,Liu2025,Gros2022,Hedrick2022,Zhao2026,Yuan2024108,Liu2024,Wu2025,Zimmermann2025,Song2022,Razmi2025,Yuan2025,Qi2024,Jiang2025,Bao2024,Fan20234763}\\\cite{Hu2024,Alfonso2024141,Liu202425024,Jardine20212595,OliveiradaSilva2025,Zuliani20255690,Ren2025,Ojand202270,Zhang20258695,He202519266,Cui20242217,Luque20241254,Wan20255353,Dai20224749,Fan2025,Khalatbarisoltani202313639,Giannini202314830,Fu20244349}\\\cite{Wang20247576,Amiri2025,Capra2025,Zhang2023,Usama2025,Li2023,MensahAkwasi2025,Zhang2022}};
			\node[align=center,font={\scriptsize}] at (0.721,0.281) {\cite{Liang2024,Peng2025,Yameen2025,Kim202542887,NejatbakhshEsfahani2024,Xie202221163,Yan2025,Locher2025,Li202512058,Xie20241507,Chen2024964,Feng20255380,Wang2023,Yang2021,Chen2024,Fang2025}};
			\node[align=center,font={\scriptsize}] at (0.721,0.231) {\cite{Jardine2019410,Jardine2021,Zhu2025,Gros2022,Hedrick2022,Liang2024,Zhao2026,Yuan2024108,Liu2024,Peng2025,Wu2025,Zimmermann2025,Song2022,Razmi2025,Yuan2025,Qi2024,Jiang2025,Bao2024}\\\cite{Yameen2025,Fan20234763,Hu2024,Kim202542887,Alfonso2024141,Liu202425024,Jardine20212595,NejatbakhshEsfahani2024,Zuliani20255690,Ren2025,Zhang20258695,He202519266,Cui20242217,Luque20241254,Wan20255353,Dai20224749,Fan2025,Xie202221163}\\\cite{Khalatbarisoltani202313639,Giannini202314830,Yan2025,Fu20244349,Wang20247576,Amiri2025,Xie20241507,Chen2024964,Feng20255380,Wang2023,Capra2025,Zhang2023,Li2023,Yang2021,Chen2024,MensahAkwasi2025,Zhang2022,Fang2025}};
			\node[align=center,font={\scriptsize}] at (0.721,0.181) {\cite{Liu2025,OliveiradaSilva2025,Ojand202270}};
			\node[align=center,font={\scriptsize}] at (0.721,0.131) {\cite{Locher2025,Li202512058,Usama2025}};
			\node[align=center,font={\scriptsize}] at (0.721,0.080) {\cite{Liang2024,Wu2025,Yameen2025,Kim202542887,Zuliani20255690,Wan20255353,Yan2025,Locher2025,Li202512058,Wang2023,Usama2025}};
			\node[align=center,font={\scriptsize}] at (0.721,0.030) {\cite{Jardine2019410,Jardine2021,Zhu2025,Liu2025,Gros2022,Hedrick2022,Zhao2026,Yuan2024108,Liu2024,Peng2025,Zimmermann2025,Song2022,Razmi2025,Yuan2025,Qi2024,Jiang2025,Bao2024,Fan20234763}\\\cite{Hu2024,Alfonso2024141,Liu202425024,Jardine20212595,NejatbakhshEsfahani2024,OliveiradaSilva2025,Ren2025,Ojand202270,Zhang20258695,He202519266,Cui20242217,Luque20241254,Dai20224749,Fan2025,Xie202221163,Khalatbarisoltani202313639,Giannini202314830,Fu20244349}\\\cite{Wang20247576,Amiri2025,Xie20241507,Chen2024964,Feng20255380,Capra2025,Zhang2023,Li2023,Yang2021,Chen2024,MensahAkwasi2025,Zhang2022,Fang2025}};
		\end{scope}
	\end{tikzpicture}
	\caption{Taxonomy of Cost Functions in MPC--RL Linear Control}
	\label{fig:taxonomy_cost_function}
\end{figure}

Summary at a Glance: Figure~\ref{fig:taxonomy_cost_function} provides a structural overview of the diversity in cost function designs across the literature. However, to understand their implementation, examining their mathematical structures is essential. Table~\ref{tab:objective_classes} synthesizes the most prevalent formulations into six distinct classes, providing their standard mathematical structures and extensive citations. Across the surveyed studies, these classes exhibit a clear distribution: quadratic tracking and least-squares formulations are the most prevalent, appearing in 40.0\% (24 out of 60) of the cases. Smoothness and rate-of-change penalties follow at 26.7\% (16 out of 60). Soft-constraint penalties and economic or operational objectives are utilized in 13.3\% (8 out of 60) and 11.7\% (7 out of 60) of the studies, respectively. Finally, mixed-norm or $\ell_1$ objectives account for 6.7\% (4 out of 60), while optimization-defined values or RL objectives represent the least frequent category at just 1.7\% (1 out of 60).

It is important to note that these six classes are overlapping representative patterns utilized within complex control structures; therefore, their percentages do not sum to 100\%.

\begin{table}[!htbp]
	\centering
	\caption{Common objective-function classes used in optimal control and MPC}
	\label{tab:objective_classes}
	\renewcommand{\arraystretch}{1.25}
	\begin{tabularx}{\textwidth}{c X X X}
		\hline
		\textbf{\#} & \textbf{Objective class} & \textbf{Description and Representative Formulation} & \textbf{Citations} \\
		\hline
		
		1 &
		Quadratic tracking / least-squares (L2) \vspace*{2mm} \[
		J = \sum_{k=0}^{N-1}\!\Big[
		(x_k-x_k^{\mathrm{ref}})^\top Q (x_k-x_k^{\mathrm{ref}})
		+ (u_k-u_k^{\mathrm{ref}})^\top R (u_k-u_k^{\mathrm{ref}})
		\Big]
		+ (x_N-x_N^{\mathrm{ref}})^\top P (x_N-x_N^{\mathrm{ref}}),
		\]&
		Weighted squared tracking errors and control effort, often with a terminal term.&
		\cite{Jardine2019410,Yuan2024108,Peng2025,Wu2025,Song2022,Yuan2025,Qi2024,Bao2024,Alfonso2024141,Liu202425024,Jardine20212595,Zuliani20255690,Ren2025,Ojand202270,Khalatbarisoltani202313639,Yan2025,Fu20244349,Locher2025,Li202512058,Chen2024964,Zhang2023,Li2023,Yang2021,Chen2024}
		\\
		\hline
		
		2 &
		Smoothness / rate-of-change penalties \vspace*{2mm}	\[
		J = \sum_{k=0}^{N-1}\Big(
		\lVert x_k-x_k^{\mathrm{ref}}\rVert_Q^2 + \lVert u_k\rVert_R^2
		\Big)
		+ \sum_{k=0}^{N-2}\lVert \Delta u_k\rVert_S^2,
		\qquad
		\Delta u_k := u_{k+1}-u_k,
		\]&
		Adds penalties on input or state variation (e.g., input rate, jerk, move suppression).&
		\cite{Jardine2021,Zhu2025,Hedrick2022,Liu2024,Razmi2025,Yameen2025,Fan20234763,Hu2024,Kim202542887,Zhang20258695,Luque20241254,Dai20224749,Fan2025,Giannini202314830,Wang20247576,Zhang2022}
		\\
		\hline
		
		3 &
		Soft-constraint penalties \vspace*{8mm} 
		\[
		\begin{aligned}
			\min_{\{u_k\},\{\epsilon_k\}} \quad
			& \sum_{k=0}^{N-1}\Big(
			\lVert x_k-x_k^{\mathrm{ref}}\rVert_Q^2 + \lVert u_k\rVert_R^2
			\Big)
			+ \rho\sum_{k=0}^{N-1}\lVert \epsilon_k\rVert_2^2 \\
			\text{s.t.}\quad
			& g(x_k,u_k) \le \epsilon_k, \qquad \epsilon_k \ge 0,
		\end{aligned}
		\],&
		Constraint violations allowed but penalized via slack variables or hinge terms.&
		\cite{Zimmermann2025,Xie202221163,Feng20255380,Usama2025,Fang2025,NejatbakhshEsfahani2024,Amiri2025,Wang2023}  \vspace*{12mm} \[\text{$\equiv$} \sum_k \alpha \max\{0,g(x_k,u_k)\}\]
		\\
		\hline
		
		4 &
		Economic / operational objectives \vspace*{5mm}
		\[
		J = \sum_{k=0}^{N-1}\Big(
		c_{\mathrm{en}} E_k
		+ c_{\mathrm{em}} \mathrm{Em}_k
		+ c_{\mathrm{deg}} D_k
		\Big)
		+ c_{\mathrm{term}}\,\phi(x_N),
		\] &
		Costs built from physical or economic metrics (energy, emissions, degradation, tariffs).&
		\cite{Liu2025,Zhao2026,Jiang2025,He202519266,Wan20255353,Capra2025,MensahAkwasi2025}
		\\
		\hline
		
		5 &
		Mixed-norm / $\ell_1$ objectives \vspace*{8mm}
		\[
		J = \sum_{k=0}^{N-1}\Big(
		\lVert x_k-x_k^{\mathrm{ref}}\rVert_Q^2
		+ \alpha \lVert u_k\rVert_1
		\Big)
		= \sum_{k=0}^{N-1}\Big(
		\lVert x_k-x_k^{\mathrm{ref}}\rVert_Q^2
		+ \alpha \sum_{i=1}^m |u_{k,i}|
		\Big),
		\] &
		Includes absolute-value or $\ell_1$-norm terms for robustness, sparsity, or linear pricing.&
		\cite{Liang2024,OliveiradaSilva2025,Cui20242217,Xie20241507}
		\\
		\hline
		
		6 &
		Optimization-defined value / RL objective \vspace*{4mm}
		\[
		Q_\theta(s,a) := \min_{z}\;
		\sum_{k=0}^{N-1} \ell_\theta(x_k,u_k)
		+ V_\theta(x_N)
		\quad \text{s.t.}\quad
		x_{k+1}=f_\theta(x_k,u_k),\;
		x_0=s,\;
		u_0=a,
		\] &
		Value or action-value defined implicitly as the optimal cost of a finite-horizon control problem.&
		\cite{Gros2022}
		\\
		\hline
		
	\end{tabularx}
\end{table}

\section{Taxonomy of RL--MPC Combinations: MPC Types}
\label{sec:mpc_types}
\begin{figure}[!htbp]
	\centering
	\begin{tikzpicture}
		\node[anchor=south west,inner sep=0] (img) at (0,0)
		{\includegraphics[width=0.95\linewidth]{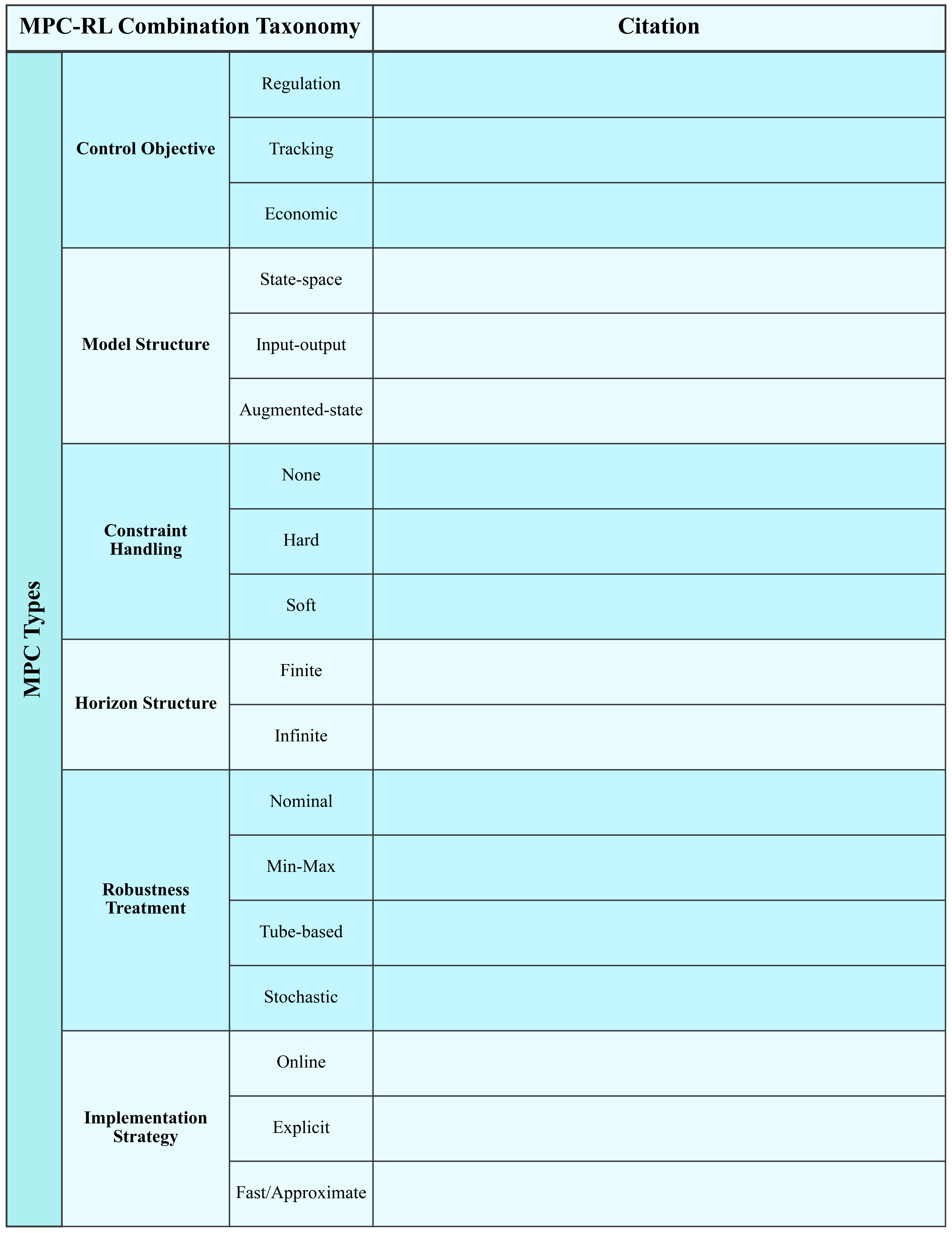}};
		\begin{scope}[x={(img.south east)},y={(img.north west)}]
			\node[align=center,font={\scriptsize}] at (0.692,0.931) {\cite{Song2022,Qi2024,Fan20234763,Zuliani20255690,Amiri2025}};
			\node[align=center,font={\scriptsize}] at (0.692,0.878) {\cite{Jardine2019410,Jardine2021,Zhu2025,Gros2022,Hedrick2022,Liang2024,Zhao2026,Yuan2024108,Liu2024,Peng2025,Wu2025,Zimmermann2025,Razmi2025,Yuan2025,Jiang2025,Bao2024,Yameen2025,Hu2024,Kim202542887,Alfonso2024141}\\\cite{Liu202425024,Jardine20212595,NejatbakhshEsfahani2024,Ren2025,Zhang20258695,He202519266,Cui20242217,Luque20241254,Wan20255353,Dai20224749,Fan2025,Xie202221163,Giannini202314830,Yan2025,Locher2025,Wang20247576,Li202512058,Xie20241507,Chen2024964,Feng20255380}\\\cite{Wang2023,Capra2025,Zhang2023,Usama2025,Li2023,Yang2021,Chen2024,MensahAkwasi2025,Zhang2022,Fang2025}};
			\node[align=center,font={\scriptsize}] at (0.692,0.825) {\cite{Liu2025,OliveiradaSilva2025,Ojand202270,Khalatbarisoltani202313639,Fu20244349}};
			\node[align=center,font={\scriptsize}] at (0.692,0.772) {\cite{Jardine2019410,Jardine2021,Liu2025,Gros2022,Hedrick2022,Zhao2026,Yuan2024108,Liu2024,Wu2025,Zimmermann2025,Song2022,Razmi2025,Yuan2025,Qi2024,Jiang2025,Yameen2025,Fan20234763,Hu2024,Kim202542887,Alfonso2024141}\\\cite{Jardine20212595,NejatbakhshEsfahani2024,OliveiradaSilva2025,Zuliani20255690,Ren2025,Ojand202270,Luque20241254,Wan20255353,Dai20224749,Fan2025,Xie202221163,Khalatbarisoltani202313639,Giannini202314830,Yan2025,Fu20244349,Locher2025,Li202512058,Amiri2025,Wang2023,Capra2025}\\\cite{Zhang2023,Usama2025,Li2023,Chen2024,MensahAkwasi2025,Zhang2022,Fang2025}};
			\node[align=center,font={\scriptsize}] at (0.692,0.720) {\cite{Zhang20258695}};
			\node[align=center,font={\scriptsize}] at (0.692,0.667) {\cite{Zhu2025,Liang2024,Peng2025,Bao2024,Liu202425024,He202519266,Cui20242217,Wang20247576,Xie20241507,Chen2024964,Feng20255380,Yang2021}};
			\node[align=center,font={\scriptsize}] at (0.692,0.614) {\cite{Cui20242217,Dai20224749}};
			\node[align=center,font={\scriptsize}] at (0.692,0.561) {\cite{Jardine2019410,Jardine2021,Zhu2025,Liu2025,Gros2022,Zhao2026,Yuan2024108,Liu2024,Wu2025,Zimmermann2025,Song2022,Razmi2025,Yuan2025,Qi2024,Jiang2025,Bao2024,Fan20234763,Hu2024,Alfonso2024141,Liu202425024}\\\cite{Jardine20212595,OliveiradaSilva2025,Ren2025,Ojand202270,Zhang20258695,He202519266,Luque20241254,Wan20255353,Fan2025,Khalatbarisoltani202313639,Giannini202314830,Fu20244349,Wang20247576,Amiri2025,Capra2025,Zhang2023,Usama2025,Li2023,MensahAkwasi2025,Zhang2022}};
			\node[align=center,font={\scriptsize}] at (0.692,0.508) {\cite{Hedrick2022,Liang2024,Peng2025,Yameen2025,Kim202542887,NejatbakhshEsfahani2024,Zuliani20255690,Xie202221163,Yan2025,Locher2025,Li202512058,Xie20241507,Chen2024964,Feng20255380,Wang2023,Yang2021,Chen2024,Fang2025}};
			\node[align=center,font={\scriptsize}] at (0.692,0.455) {\cite{Jardine2019410,Jardine2021,Zhu2025,Liu2025,Gros2022,Hedrick2022,Liang2024,Zhao2026,Yuan2024108,Liu2024,Peng2025,Wu2025,Zimmermann2025,Razmi2025,Yuan2025,Qi2024,Jiang2025,Bao2024,Yameen2025,Fan20234763}\\\cite{Hu2024,Kim202542887,Alfonso2024141,Liu202425024,Jardine20212595,NejatbakhshEsfahani2024,OliveiradaSilva2025,Zuliani20255690,Ren2025,Ojand202270,Zhang20258695,He202519266,Cui20242217,Luque20241254,Wan20255353,Dai20224749,Fan2025,Xie202221163,Khalatbarisoltani202313639,Giannini202314830}\\\cite{Yan2025,Fu20244349,Locher2025,Wang20247576,Li202512058,Amiri2025,Xie20241507,Chen2024964,Feng20255380,Wang2023,Capra2025,Zhang2023,Usama2025,Li2023,Yang2021,Chen2024,MensahAkwasi2025,Zhang2022,Fang2025}};
			\node[align=center,font={\scriptsize}] at (0.692,0.402) {\cite{Song2022}};
			\node[align=center,font={\scriptsize}] at (0.692,0.349) {\cite{Jardine2019410,Jardine2021,Zhu2025,Liu2025,Hedrick2022,Liang2024,Zhao2026,Yuan2024108,Liu2024,Peng2025,Wu2025,Zimmermann2025,Razmi2025,Yuan2025,Qi2024,Jiang2025,Bao2024,Yameen2025,Fan20234763,Hu2024}\\\cite{Kim202542887,Alfonso2024141,Liu202425024,Jardine20212595,NejatbakhshEsfahani2024,OliveiradaSilva2025,Zuliani20255690,Ren2025,Ojand202270,Zhang20258695,He202519266,Cui20242217,Luque20241254,Wan20255353,Dai20224749,Fan2025,Xie202221163,Khalatbarisoltani202313639,Giannini202314830,Yan2025}\\\cite{Fu20244349,Wang20247576,Li202512058,Amiri2025,Xie20241507,Chen2024964,Feng20255380,Wang2023,Capra2025,Zhang2023,Usama2025,Li2023,Yang2021,Chen2024,MensahAkwasi2025,Zhang2022,Fang2025}};
			\node[align=center,font={\scriptsize}] at (0.692,0.296) {\cite{Song2022}};
			\node[align=center,font={\scriptsize}] at (0.692,0.243) {\cite{Gros2022}};
			\node[align=center,font={\scriptsize}] at (0.692,0.190) {\cite{Locher2025}};
			\node[align=center,font={\scriptsize}] at (0.692,0.138) {\cite{Jardine2019410,Jardine2021,Zhu2025,Liu2025,Gros2022,Hedrick2022,Liang2024,Zhao2026,Yuan2024108,Liu2024,Peng2025,Wu2025,Zimmermann2025,Razmi2025,Yuan2025,Jiang2025,Bao2024,Yameen2025,Fan20234763,Hu2024}\\\cite{Kim202542887,Alfonso2024141,Liu202425024,Jardine20212595,NejatbakhshEsfahani2024,Zuliani20255690,Ren2025,Ojand202270,Zhang20258695,He202519266,Cui20242217,Luque20241254,Wan20255353,Dai20224749,Fan2025,Xie202221163,Khalatbarisoltani202313639,Giannini202314830,Yan2025,Fu20244349}\\\cite{Wang20247576,Li202512058,Amiri2025,Xie20241507,Chen2024964,Feng20255380,Wang2023,Capra2025,Zhang2023,Usama2025,Li2023,Yang2021,Chen2024,MensahAkwasi2025,Zhang2022,Fang2025}};
			\node[align=center,font={\scriptsize}] at (0.692,0.085) {\cite{Song2022}};
			\node[align=center,font={\scriptsize}] at (0.692,0.032) {\cite{Qi2024,OliveiradaSilva2025,Locher2025}};
		\end{scope}
	\end{tikzpicture}
	\caption{Taxonomy of MPC Formulations}
	\label{Taxonomy_mpc}
\end{figure}

Linear MPC is not a single, monolithic control law, but rather a versatile framework encompassing a wide spectrum of controllers tailored to specific application requirements within linear or linearized predictive control settings. These controllers can be systematically categorized along six primary dimensions: control objective, model structure, constraint handling, horizon structure, robustness treatment, and implementation strategy. This multidimensional design space is illustrated in Figure \ref{Taxonomy_mpc}. 
 In the context of the reviewed RL--MPC literature, understanding these MPC types is crucial, as the characteristics of the chosen MPC formulation influence the degree of freedom, feasibility, and specific role that RL can assume within the hybrid architecture.

\subsection{Classification by Control Objective}
The fundamental goal of the controller defines its cost function structure, which in turn influences the mathematical properties of the underlying optimization problem:

\begin{itemize}
    \item \textbf{Regulation MPC:} Designed to stabilize a system around a fixed equilibrium point or origin. It minimizes the deviations of states and inputs from this equilibrium. Deeply rooted in finite-horizon Linear Quadratic Regulator (LQR) theory, it is primarily employed for stabilization and disturbance rejection.
    \item \textbf{Tracking MPC:} Formulated to follow a dynamic reference trajectory or time-varying setpoints. It penalizes deviations from predicted states or outputs to these references and is the industry standard for motion control, robotics, and automotive applications.
    \item \textbf{Economic MPC (EMPC):} Abandons the traditional setpoint-tracking paradigm to directly optimize an economic or operational criterion (e.g., profit maximization, energy minimization, or resource efficiency). The objective function is often non-quadratic and potentially non-convex, requiring specialized stability analysis frameworks.
\end{itemize}

\textbf{RL Integration Perspective:} In the reviewed MPC--RL frameworks, the control objective defines the intervention boundary for RL. For Tracking MPC, RL is often employed as a high-level reference governor, adapting setpoints dynamically. Conversely, in Economic MPC, RL is utilized for reward shaping and cost augmentation, often approximating an economic value function while the MPC handles operational constraints.

\subsection{Classification by Model Structure}
The internal prediction model governs how the controller anticipates future behavior and defines the structure of the optimization variables:

\begin{itemize}
    \item \textbf{State-Space Models:} Expressed as $x_{k+1} = Ax_k + Bu_k$, this formulation provides a complete internal representation of the system dynamics and is the dominant structure in modern control theory due to its compatibility with rigorous stability proofs and Lyapunov-based analysis.
    \item \textbf{Input-Output Models:} Utilizing transfer functions or step-response models (e.g., Dynamic Matrix Control (DMC) or Generalized Predictive Control (GPC)). These are highly prevalent in process industries due to the ease of empirical identification from plant data.
    \item \textbf{Augmented-State Models:} Incorporate disturbance models or integrator states into the nominal model to achieve offset-free tracking and robust disturbance rejection in the presence of unmeasured constant disturbances or plant-model mismatch.
\end{itemize}

\textbf{RL Integration Perspective:} In the reviewed studies, RL interacts with the model structure through system identification or adaptive mechanisms within linear or linearized predictive models. 
 For state-space models, RL agents may adapt model-related quantities, estimate unmeasurable states, or compensate for model mismatch around nominal  $A$ and $B$ matrices.
 When using augmented states, RL can dynamically adjust disturbance-estimation terms based on the operational regime, helping reduce the gap between nominal models and real-world uncertainties.
\subsection{Classification by Constraint Handling}
One of the defining features of MPC is its explicit handling of operational limits. The strictness of these limits shapes the optimization geometry:

\begin{itemize}
    \item \textbf{None (Unconstrained MPC):} Assumes infinite control authority and boundless state spaces, often yielding closed-form analytical solutions similar to unconstrained LQR.
    \item \textbf{Hard-Constrained MPC:} Enforces strict, non-relaxable limits on states and inputs. While essential for safety-critical applications, it risks optimization infeasibility if disturbances push the system outside the maximum control invariant set.
    \item \textbf{Soft-Constrained MPC:} Introduces slack variables to state and output constraints (input constraints remain hard due to physical actuator limits). This approach penalizes constraint violations rather than forbidding them, significantly enhancing algorithmic stability and preventing solver failure during transient excursions.
\end{itemize}

\textbf{RL Integration Perspective:} Safe RL architectures often rely on the constraint handling capabilities of MPC. RL is often tasked with dynamically tuning the penalty weights of slack variables in Soft-Constrained MPC, balancing performance against the severity of constraint violation. In Hard-Constrained setups, the MPC often acts as an absolute safety filter for exploratory RL actions.

\subsection{Classification by Horizon Structure}
The temporal depth of the prediction and control windows dictates the trade-off between closed-loop optimality and real-time computational feasibility:

\begin{itemize}
    \item \textbf{Finite-Horizon MPC:} Utilizes a truncated prediction window. To guarantee nominal stability, it often requires the inclusion of a terminal cost and a terminal constraint set, serving as a Lyapunov function surrogate.
    \item \textbf{Infinite-Horizon MPC:} Aims to optimize performance over an infinite horizon, guaranteeing nominal stability. In practice, it is recast into a finite-horizon problem using algebraic Riccati equations to compute the exact terminal penalty.
\end{itemize}

\textbf{RL Integration Perspective:} RL has shown promise in mitigating the computational burden of long horizons. By utilizing Value Function Approximation (VFA), an RL agent can learn the approximate terminal cost offline. This allows a short, real-time feasible Finite-Horizon MPC to approximate some of the performance benefits typically associated with an Infinite-Horizon controller.

\subsection{Classification by Robustness Treatment}
The methodology used to address plant-model mismatch and external disturbances defines the controller's robustness profile:

\begin{itemize}
    \item \textbf{Nominal MPC:} Assumes a perfect deterministic model. It relies solely on the inherent robustness of the receding horizon feedback loop, which may fail under severe uncertainties.
    \item \textbf{Min-Max (Robust) MPC:} Optimizes for the worst-case scenario within a bounded uncertainty set. While ensuring strict constraint satisfaction, it is often overly conservative and computationally demanding.
    \item \textbf{Tube-Based MPC:} Decouples the problem into a nominal MPC trajectory and an ancillary feedback controller that keeps the actual system within a bounded ``tube'' around the nominal path. It strikes a sophisticated balance between performance and robustness by tightening constraints.
    \item \textbf{Stochastic MPC:} Handles uncertainties described by probability distributions, employing chance constraints to allow a pre-defined probability of constraint violation, thereby reducing the conservatism of purely robust approaches.
\end{itemize}

\textbf{RL Integration Perspective:} RL can enhance robustness treatments by complementing conservative worst-case assumptions with learned probabilistic or adaptive behaviors. In Tube-Based MPC, some studies use RL to adapt robustness-related quantities based on the current environment. Furthermore, RL can learn residual dynamic models to compensate for nominal model deficiencies, enabling a more adaptive robustness strategy within linear or linearized predictive control settings.

\subsection{Classification by Implementation Strategy}
The execution paradigm dictates the hardware requirements and the feasible sampling rates of the controller:

\begin{itemize}
    \item \textbf{Online Optimization MPC:} Solves the full constrained Quadratic Program (QP) or Non-Linear Program (NLP) iteratively at every sampling instant.
    \item \textbf{Explicit MPC:} Moves the computational burden offline by using multiparametric programming. The state space is partitioned into polyhedral regions, and the control law is stored as a Piecewise Affine (PWA) function, allowing for microsecond execution times via simple table look-ups.
    \item \textbf{Fast/Approximate MPC:} Utilizes suboptimal solvers, early-termination algorithms, or neural network approximations to compute a near-optimal control action within a strict time budget.
\end{itemize}

\textbf{RL Integration Perspective:} The implementation strategy strongly influences the viability of RL integration. RL is often synergistic with Online and Fast/Approximate MPC, where the agent can update cost function parameters or constraints dynamically. Conversely, Explicit MPC may limit RL applicability, because the PWA control law is pre-compiled offline and therefore offers less flexibility for dynamic online adaptation through RL.

\section{Taxonomy of RL--MPC Combinations: Applications}
\label{sec:applications}

\begin{figure}[!htbp]
	\centering
	\begin{tikzpicture}
		\node[anchor=south west,inner sep=0] (img) at (0,0)
		{\includegraphics[width=0.7\linewidth]{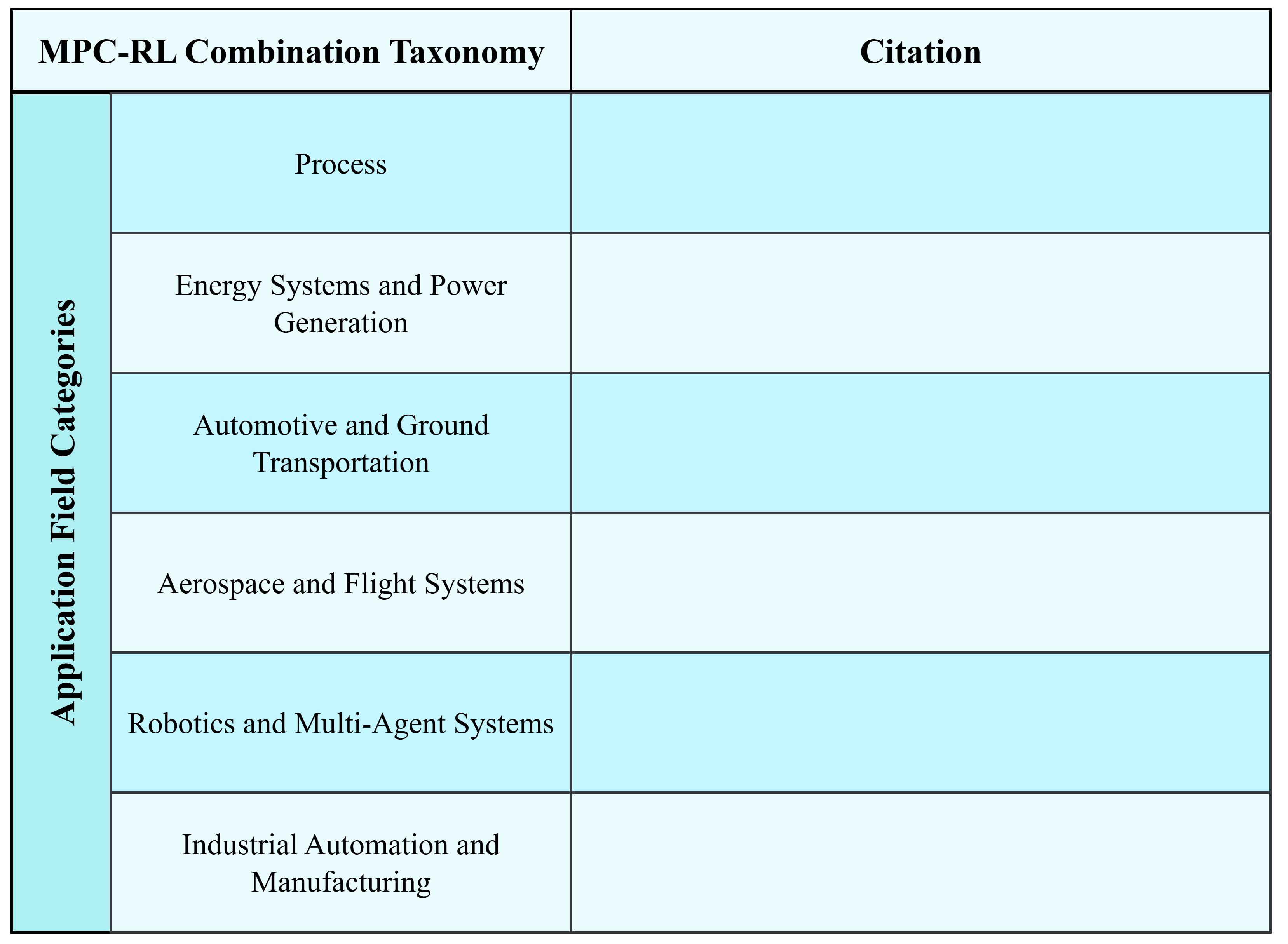}};
		\begin{scope}[x={(img.south east)},y={(img.north west)}]
			\node[align=center,font={\scriptsize}] at (0.733,0.802) {\cite{Liu2025,Hedrick2022,Zimmermann2025,Jiang2025,NejatbakhshEsfahani2024,Ojand202270,Zhang20258695,He202519266,Dai20224749}};
			\node[align=center,font={\scriptsize}] at (0.733,0.659) {\cite{Zhu2025,Liu2025,Hedrick2022,Song2022,Razmi2025,Yuan2025,Jiang2025,Yameen2025,Fan20234763,OliveiradaSilva2025,Ojand202270,Cui20242217}\\\cite{Wan20255353,Khalatbarisoltani202313639,Fu20244349,Locher2025,Amiri2025,Usama2025,MensahAkwasi2025}};
			\node[align=center,font={\scriptsize}] at (0.733,0.516) {\cite{Jardine2019410,Liang2024,Zhao2026,Yuan2024108,Liu2024,Peng2025,Wu2025,Yuan2025,Fan20234763,Kim202542887,Alfonso2024141,Ojand202270}\\\cite{Fan2025,Xie202221163,Khalatbarisoltani202313639,Giannini202314830,Yan2025,Wang20247576,Li202512058,Xie20241507,Chen2024964,Feng20255380,Wang2023,Usama2025}\\\cite{Yang2021,Chen2024,Zhang2022,Fang2025}};
			\node[align=center,font={\scriptsize}] at (0.733,0.373) {\cite{Jardine2021,Hu2024,Jardine20212595,Ren2025,Capra2025,Usama2025}};
			\node[align=center,font={\scriptsize}] at (0.733,0.230) {\cite{Jardine2019410,Jardine2021,Yuan2024108,Liu2024,Wu2025,Razmi2025,Bao2024,Fan20234763,Hu2024,Kim202542887,Alfonso2024141,Liu202425024}\\\cite{Jardine20212595,Zuliani20255690,Ren2025,Luque20241254,Fan2025,Xie202221163,Khalatbarisoltani202313639,Giannini202314830,Yan2025,Fu20244349,Chen2024964,Capra2025}\\\cite{Zhang2023,Li2023,Yang2021}};
			\node[align=center,font={\scriptsize}] at (0.733,0.087) {\cite{Liu2024,Bao2024,Liu202425024,Zuliani20255690,Zhang20258695,He202519266,Luque20241254,Dai20224749,Usama2025,Li2023}};
		\end{scope}
	\end{tikzpicture}
	\caption{Taxonomy of MPC--RL application fields in linear control}
	\label{taxonomy_application}
\end{figure}

The integration of RL with MPC has transcended theoretical boundaries, proving highly effective across a diverse array of engineering and socioeconomic domains. The inherent ability of this hybrid architecture to handle complex constraints while dynamically adapting to uncertainties makes it a versatile tool for modern control challenges. As illustrated in Figure \ref{taxonomy_application} and its condensed counterpart, the application landscape of RL--MPC integrations can be broadly categorized into the following ten principal domains:

\begin{itemize}
    \item \textbf{Process:} This overarching category includes traditional applications such as chemical reactors, and distillation columns. For the purpose of this review, it broadly encompasses Environmental and Water Systems (e.g., wastewater treatment and irrigation), Buildings and HVAC Systems (thermal comfort and climate control), and Biomedical and Healthcare Systems (such as artificial pancreas systems). Across these domains, handling plant nonlinearities, time delays, and strict operational constraints remains a prominent shared challenge, although the included studies employ linear or linearized predictive models within the MPC layer.
    
    \item \textbf{Energy Systems and Power Generation:} Encompassing microgrids, renewable energy integration, and smart grids that require robust, real-time demand-response optimization under high uncertainty.
    
    \item \textbf{Automotive and Ground Transportation:} Featuring autonomous driving, advanced driver-assistance systems (ADAS), and energy management in hybrid/electric vehicles. Within this taxonomy, this category is structurally extended to incorporate Economics, Finance, and Resource Management applications (such as supply chain logistics and portfolio management), as they often share similar underlying predictive routing and dynamic resource allocation characteristics.
    
    \item \textbf{Aerospace and Flight Systems:} Involving unmanned aerial vehicles (UAVs), spacecraft attitude control, and trajectory optimization operating under severe aerodynamic uncertainties and fast-scale dynamics.
    
    \item \textbf{Robotics and Multi-Agent Systems:} Covering robotic manipulators, legged locomotion, and swarm robotics, where high-dimensional continuous action spaces meet strict safety and collision-avoidance boundaries.
    
    \item \textbf{Industrial Automation and Manufacturing:} Addressing factory floor automation, robotic assembly lines, and predictive maintenance to ensure high-throughput and reliable production environments.
\end{itemize}

\begin{figure}[!htbp]
	\centering
	\includegraphics[width=0.65\linewidth]{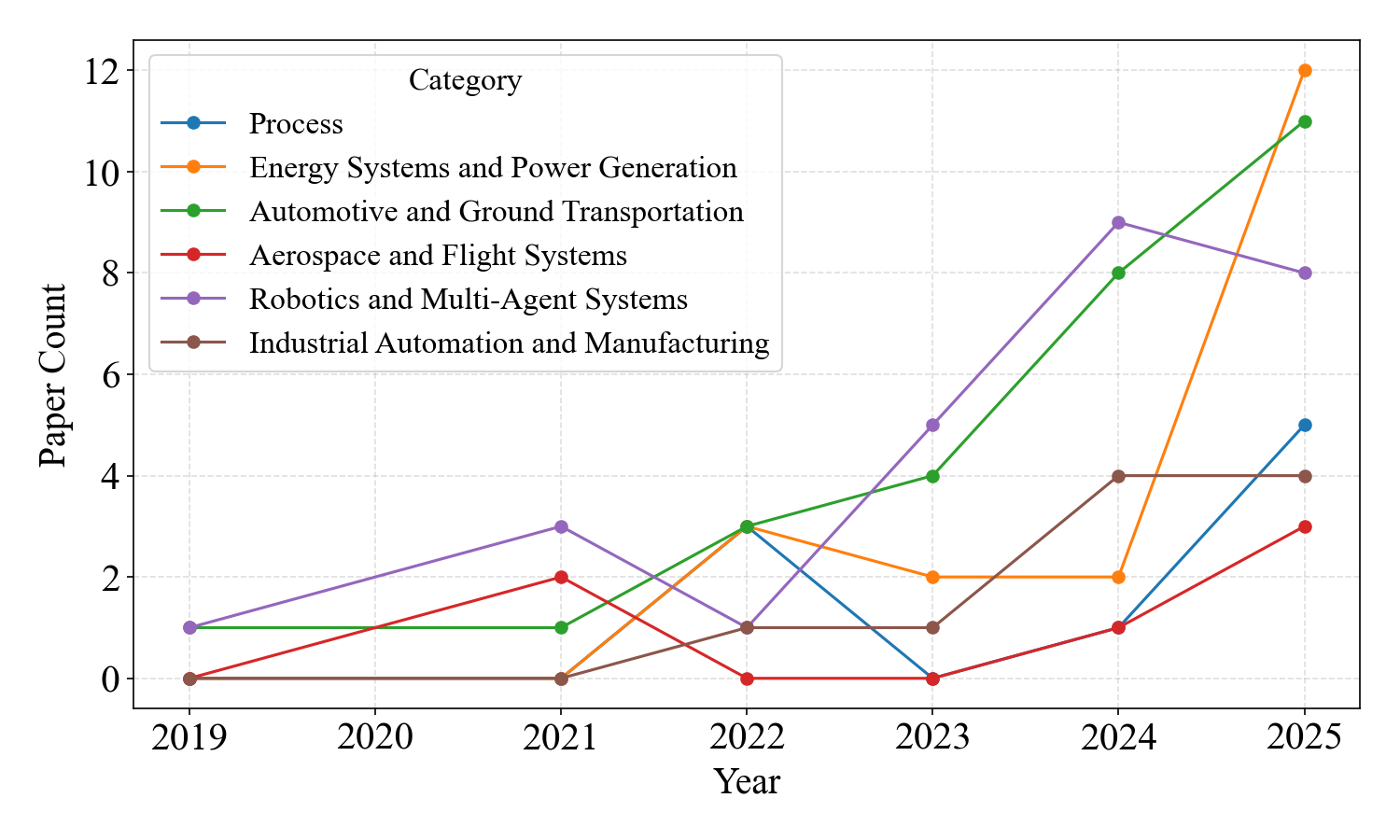}
	\caption{Application fields of MPC--RL integration during years}
	\label{Application_Year}
\end{figure}

The temporal evolution of these applications is depicted in Figure \ref{Application_Year}, which highlights the growing research interest and paradigm shift towards hybrid control architectures over recent years. As computational resources have become more accessible, domains such as automotive and robotics have seen an exponential rise in RL--MPC adoptions.


\section{Relation Between the Taxonomies: A Cross-Dimensional Analysis}
\label{sec:relations}

The true structural insights of this survey emerge not from isolating the constituent taxonomies, but from analyzing their mutual intersections. Having independently classified the RL roles, RL algorithmic frameworks, cost function structures, MPC architectures, and application domains, this section synthesizes these dimensions to reveal underlying patterns, dominant design paradigms, and synergistic mappings in the current literature.

\begin{figure}[!htbp]
	\centering
	\includegraphics[width=0.7\linewidth]{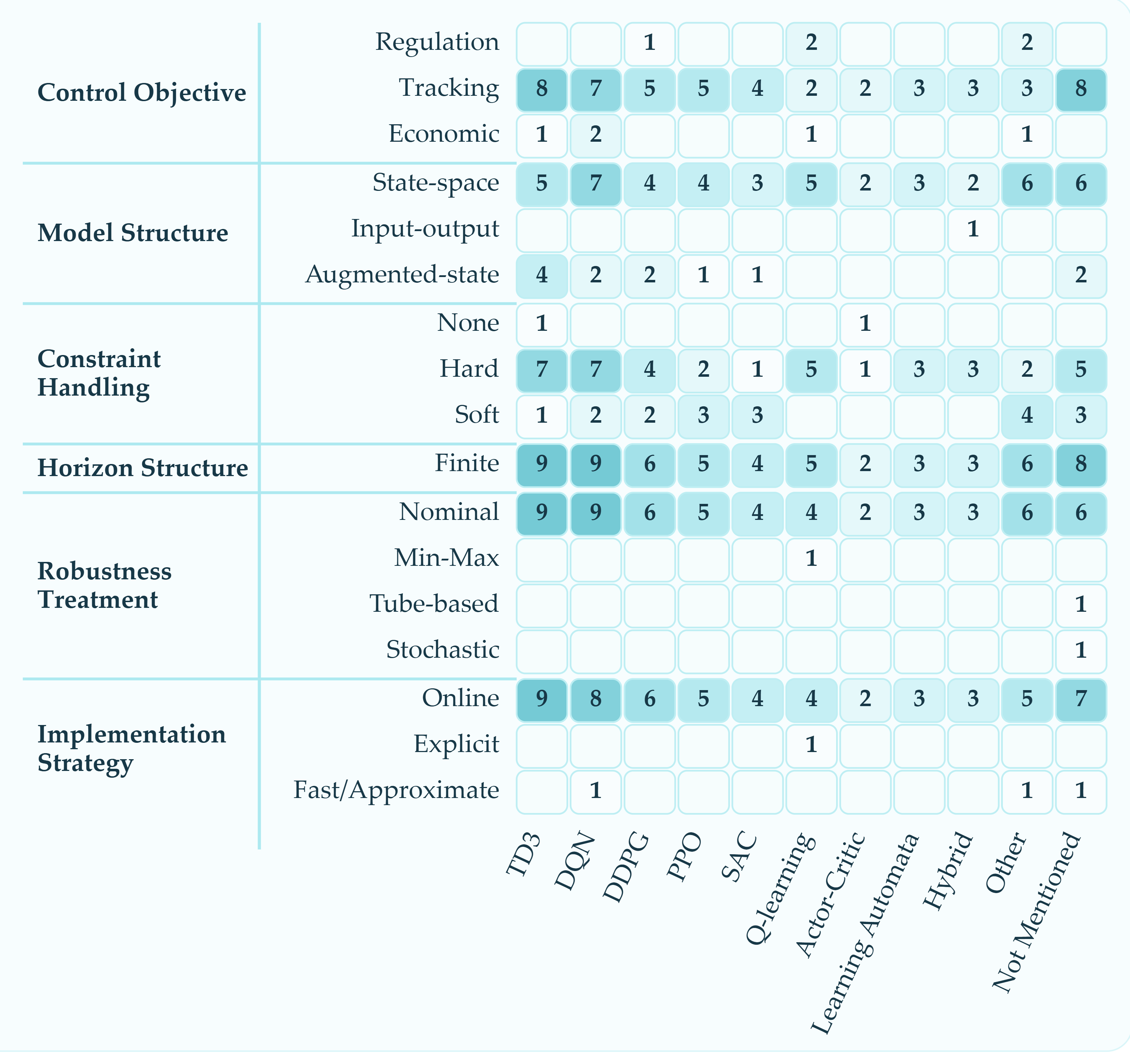}
	\caption{MPC type vs. RL algorithms in the integration of MPC--RL for linear control}
	\label{Tax2D_MPC_RL_Type}
\end{figure}

\begin{figure}[!htbp]
	\centering
	\includegraphics[width=0.7\linewidth]{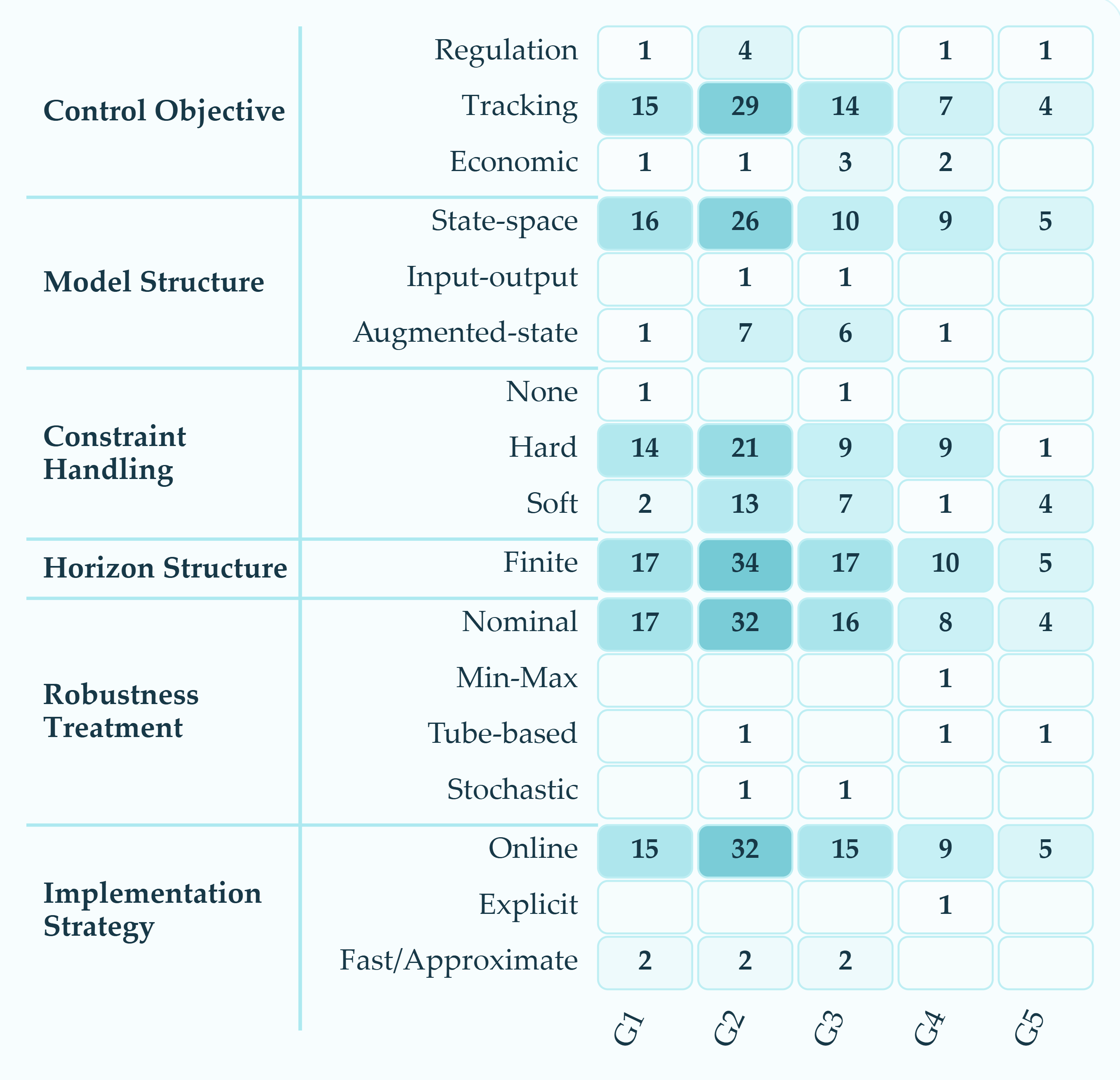}
	\caption{MPC type vs. RL roles in the integration of MPC--RL for linear control}
	\label{Tax2D_MPC_RL_Role}
\end{figure}

\begin{figure}[!htbp]
	\centering
	\includegraphics[width=0.5\linewidth]{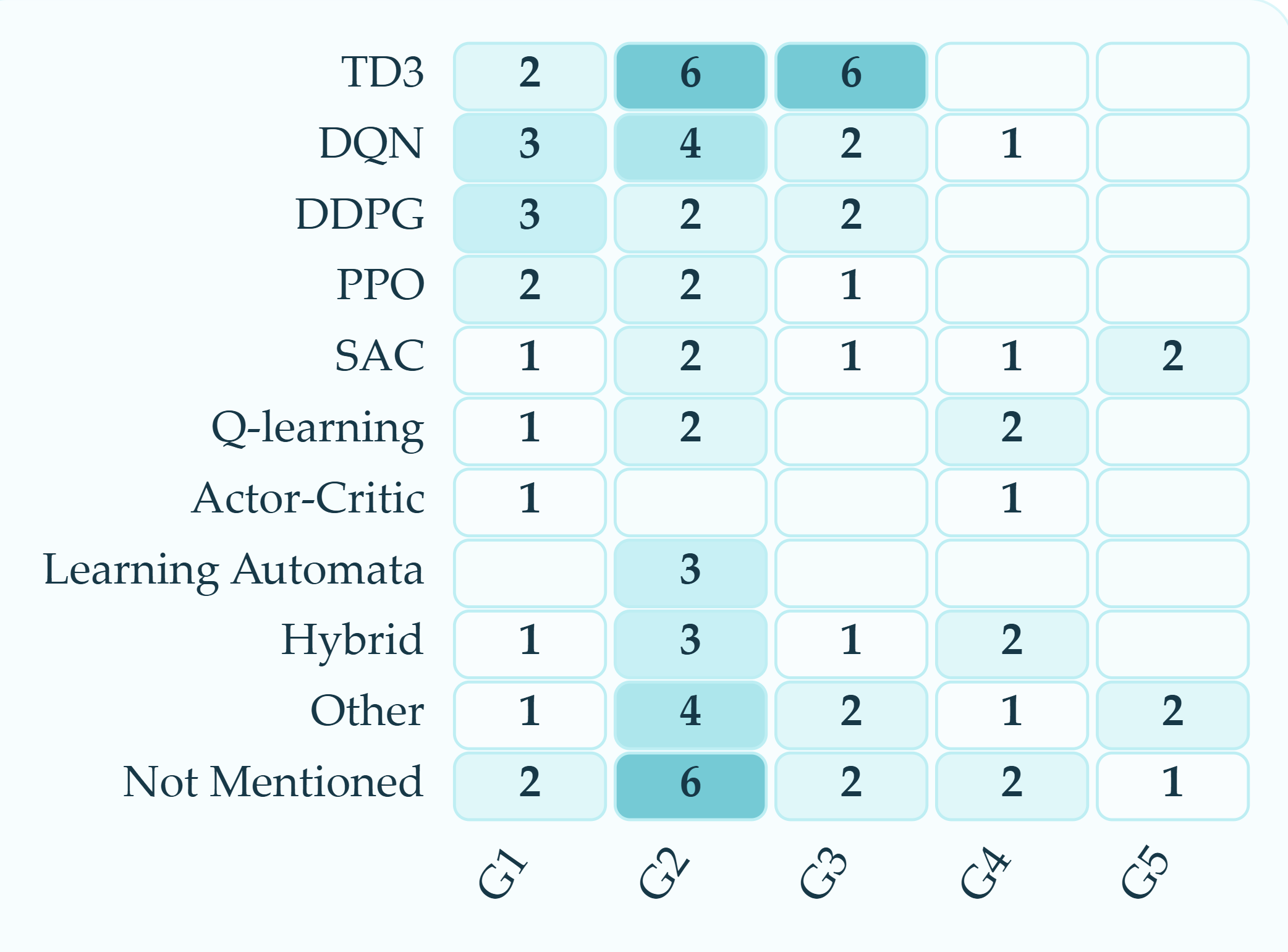}
	\caption{RL algorithms vs. RL roles in the integration of MPC--RL for linear control}
	\label{Tax2D_RL_Role_RL_Type}
\end{figure}

\begin{figure}[!htbp]
	\centering
	\includegraphics[width=0.5\linewidth]{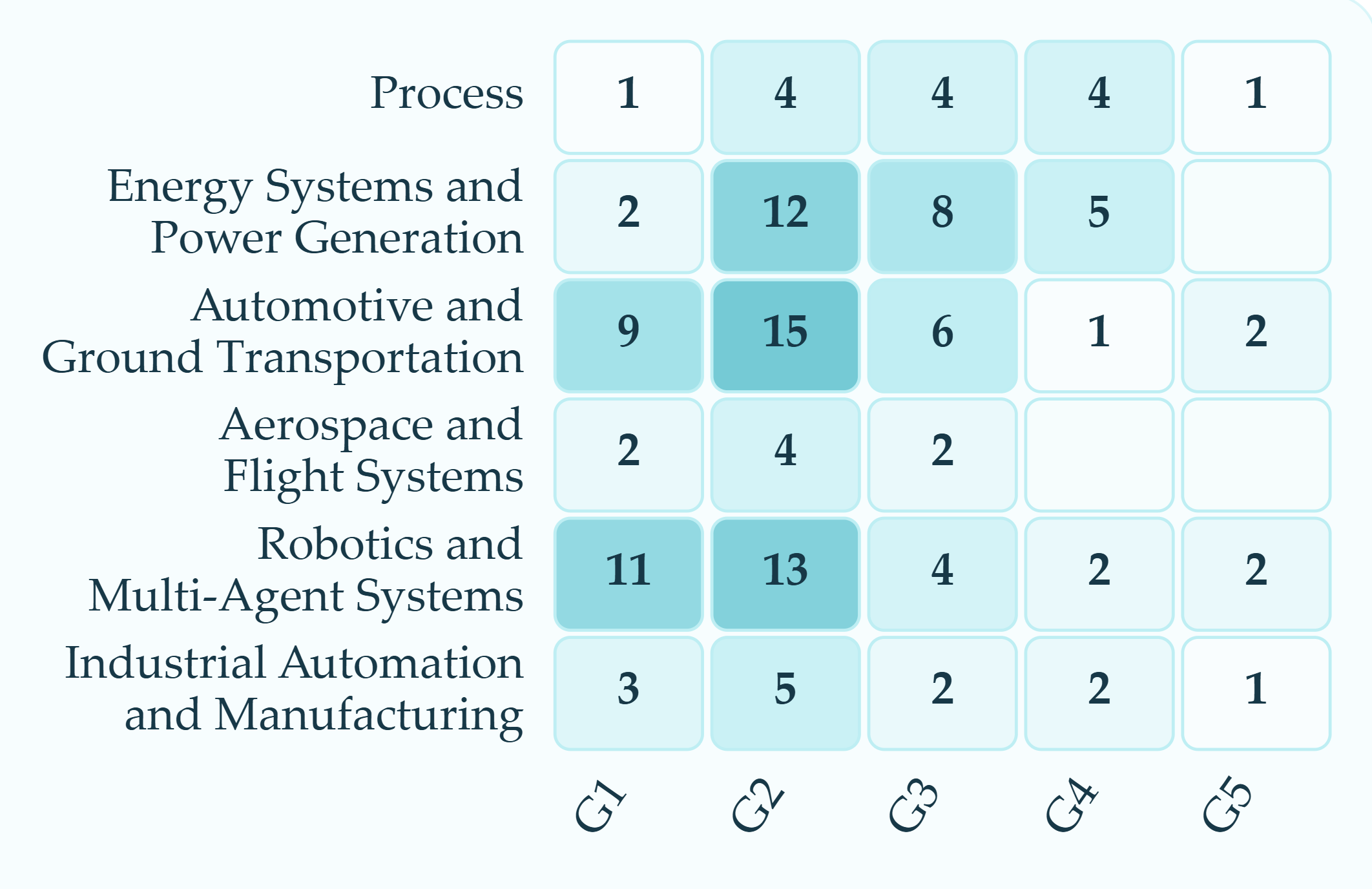}
	\caption{Application fields vs. RL roles in the integration of MPC--RL for linear control}
	\label{Tax2D_RL_Role_Applications}
\end{figure}

\begin{figure}[!htbp]
	\centering
	\includegraphics[width=0.65\linewidth]{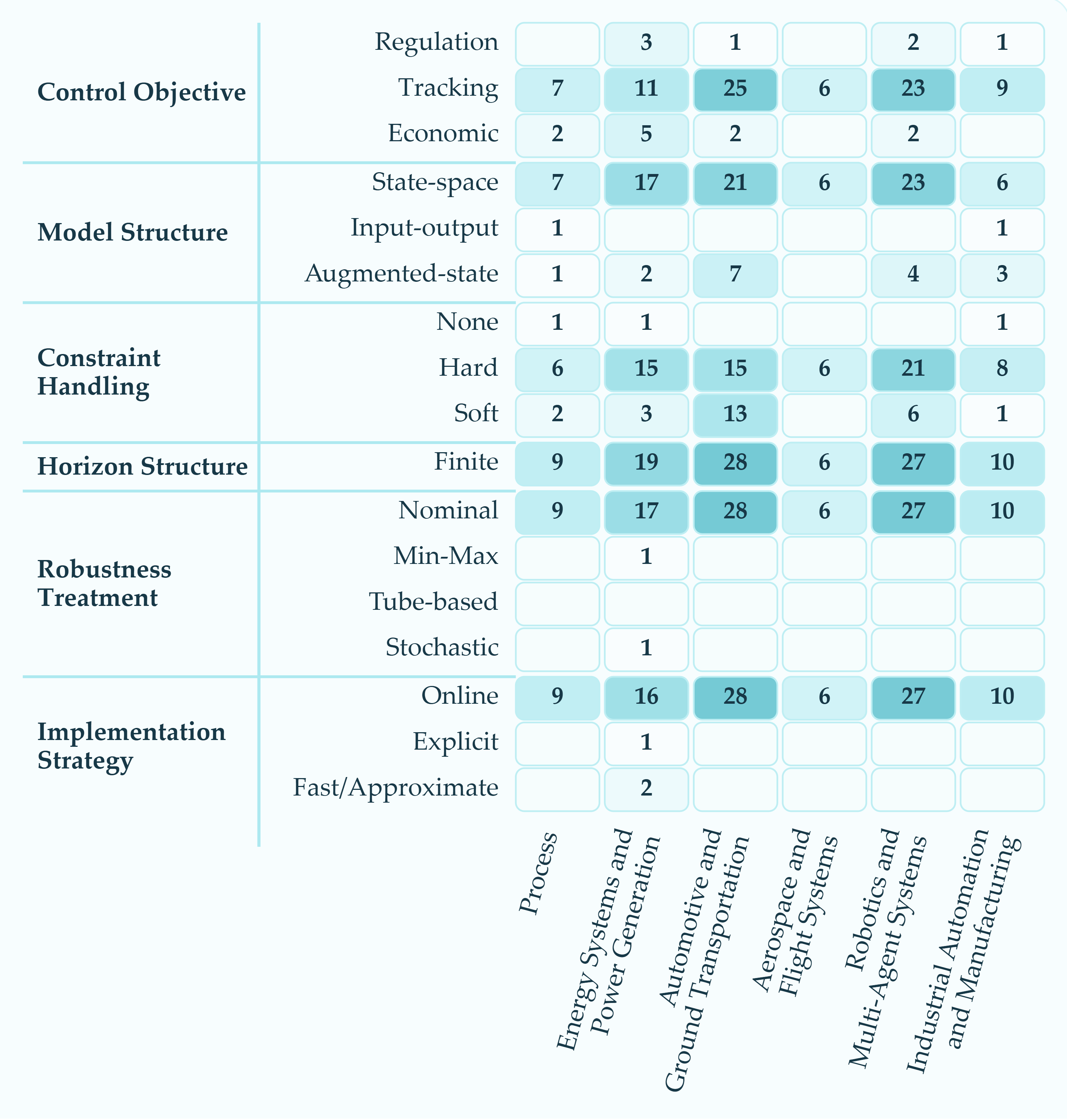}
	\caption{MPC type vs. application fields in the integration of MPC--RL for linear control}
	\label{Tax2D_MPC_Applications}
\end{figure}

\begin{figure}[!htbp]
	\centering
	\includegraphics[width=0.55\linewidth]{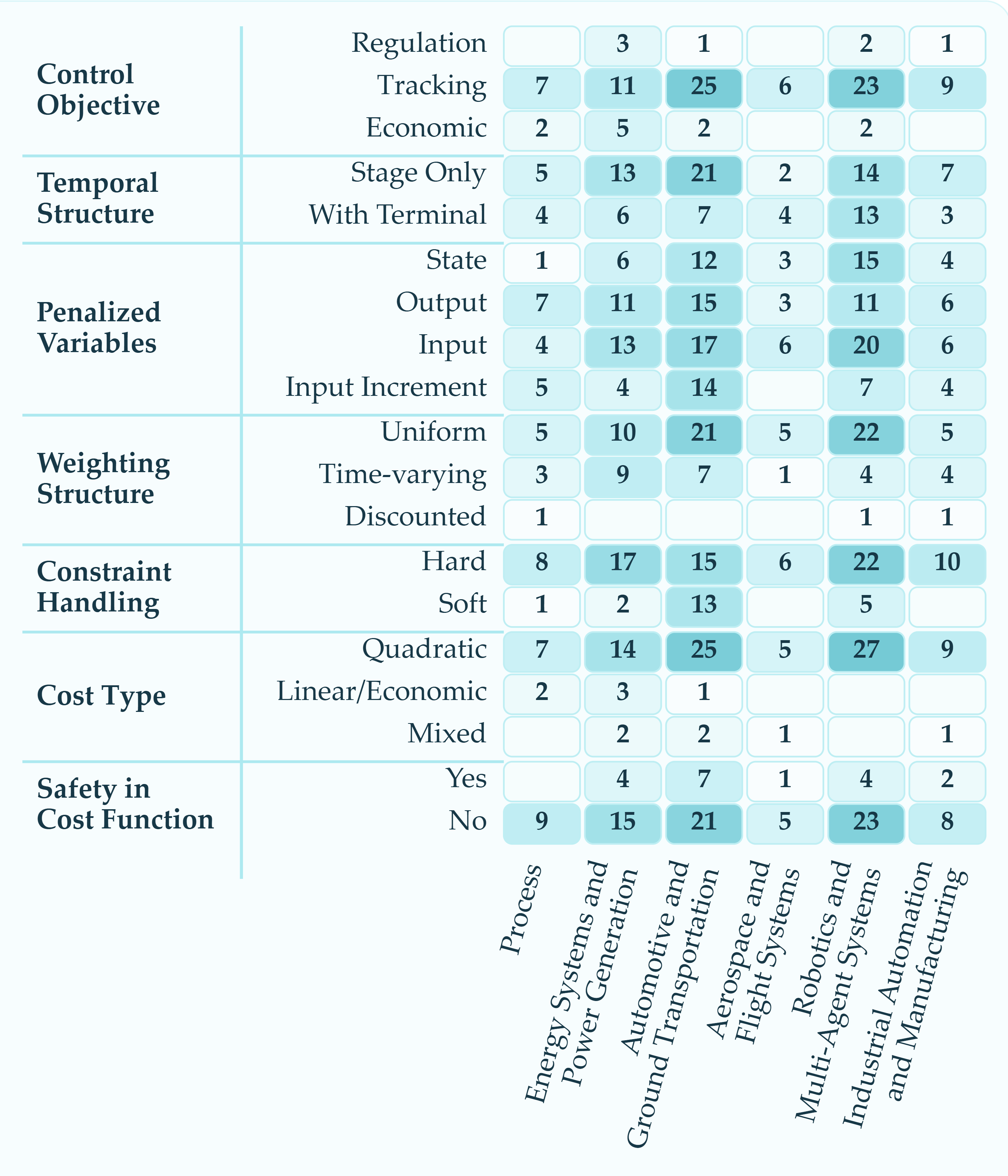}
	\caption{MPC cost functions vs. application fields in the integration of MPC--RL for linear control}
	\label{Tax2D_Cost_Applications}
\end{figure}

\begin{figure}[!htbp]
	\centering
	\includegraphics[width=0.85\linewidth]{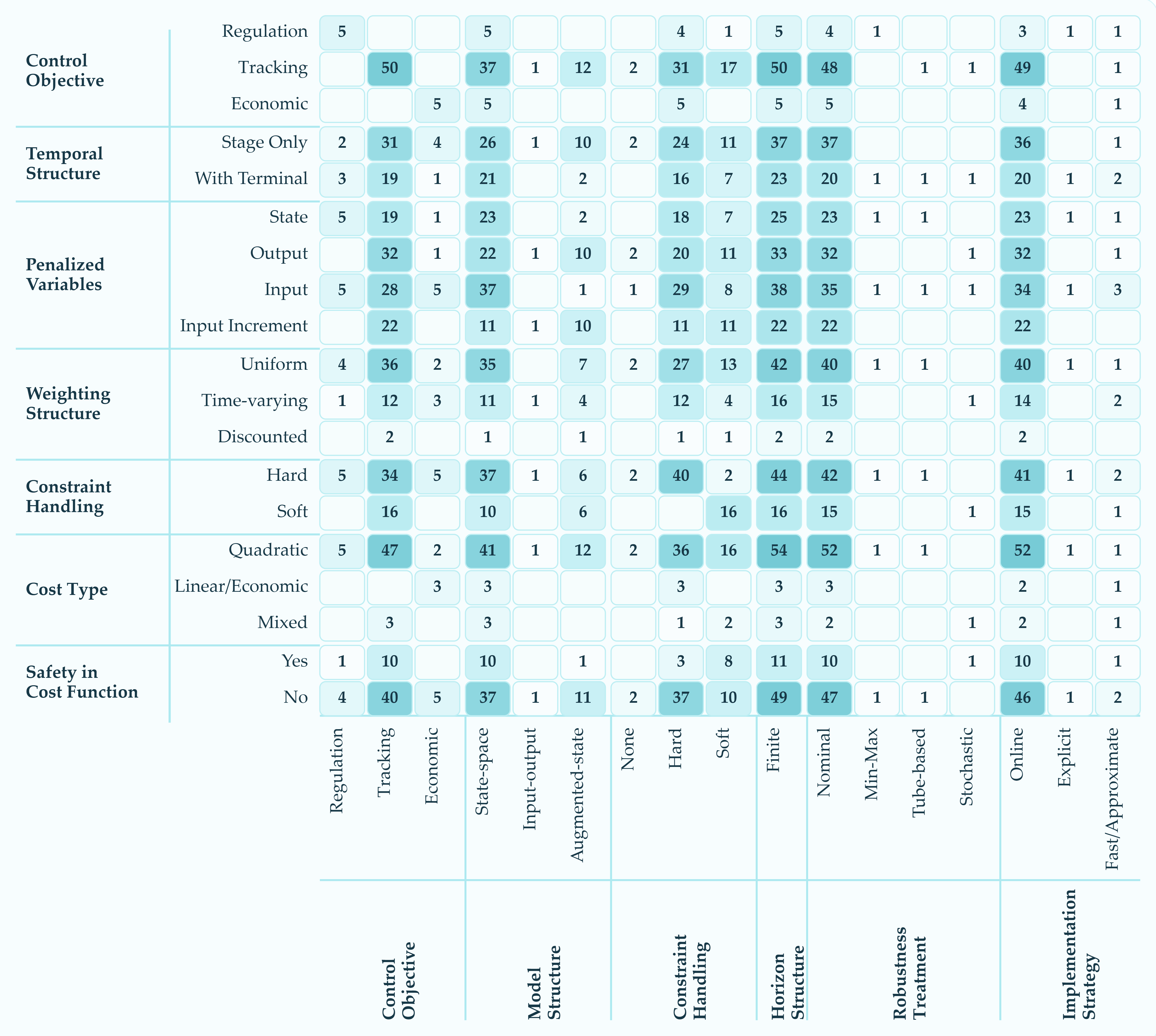}
	\caption{MPC cost functions vs. MPC type in the integration of MPC--RL for linear control}
	\label{Tax2D_Cost_MPC}
\end{figure}

\begin{figure}[!htbp]
	\centering
	\includegraphics[width=0.65\linewidth]{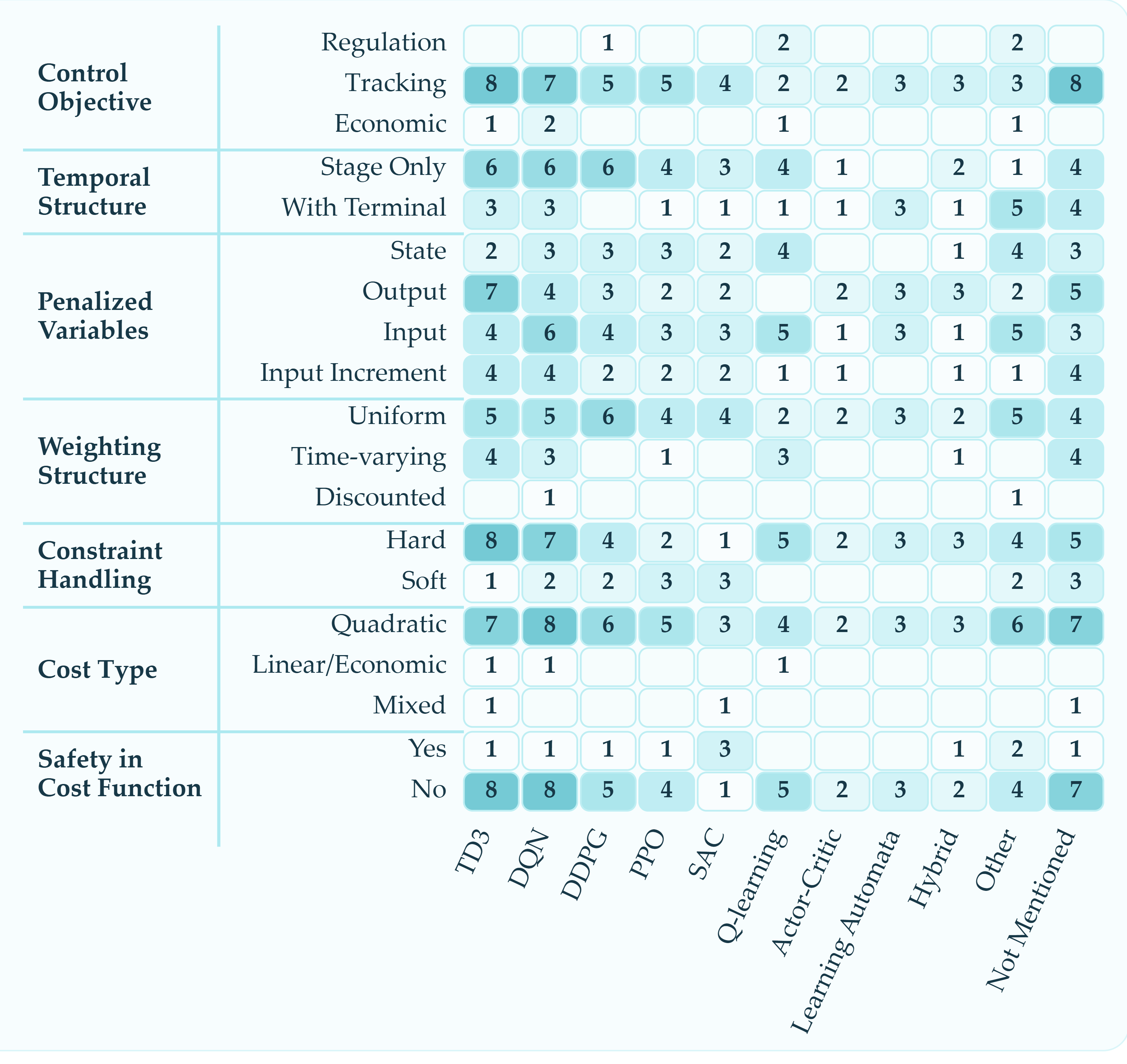}
	\caption{MPC cost functions vs. RL algorithms in the integration of MPC--RL for linear control}
	\label{Tax2D_Cost_RL_Types}
\end{figure}

\begin{figure}[!htbp]
	\centering
	\includegraphics[width=0.65\linewidth]{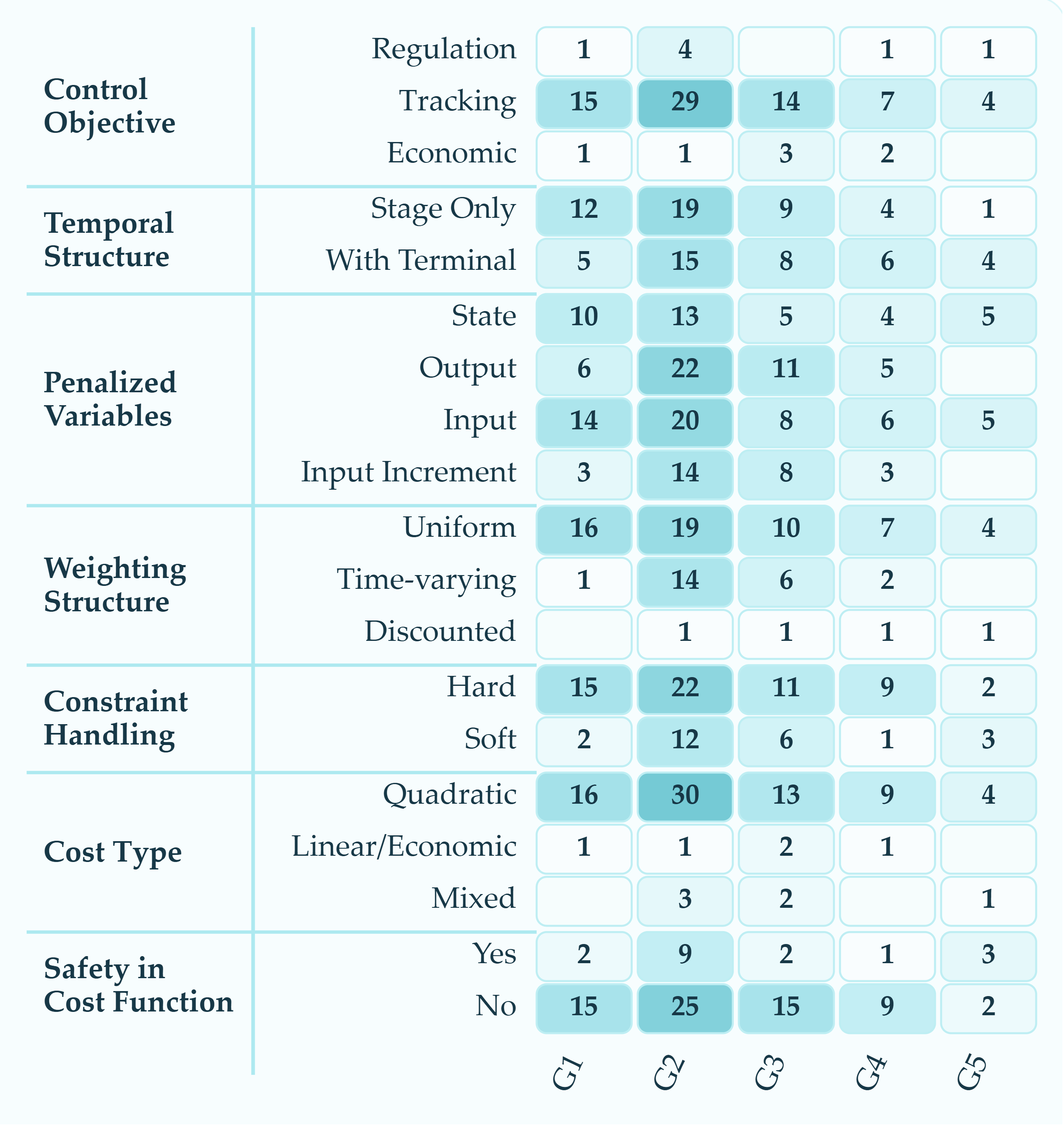}
	\caption{MPC cost functions vs. RL roles in the integration of MPC--RL for linear control}
	\label{Tax2D_Cost_RL_Roles}
\end{figure}

\begin{figure}[!htbp]
	\centering
	\includegraphics[width=0.55\linewidth]{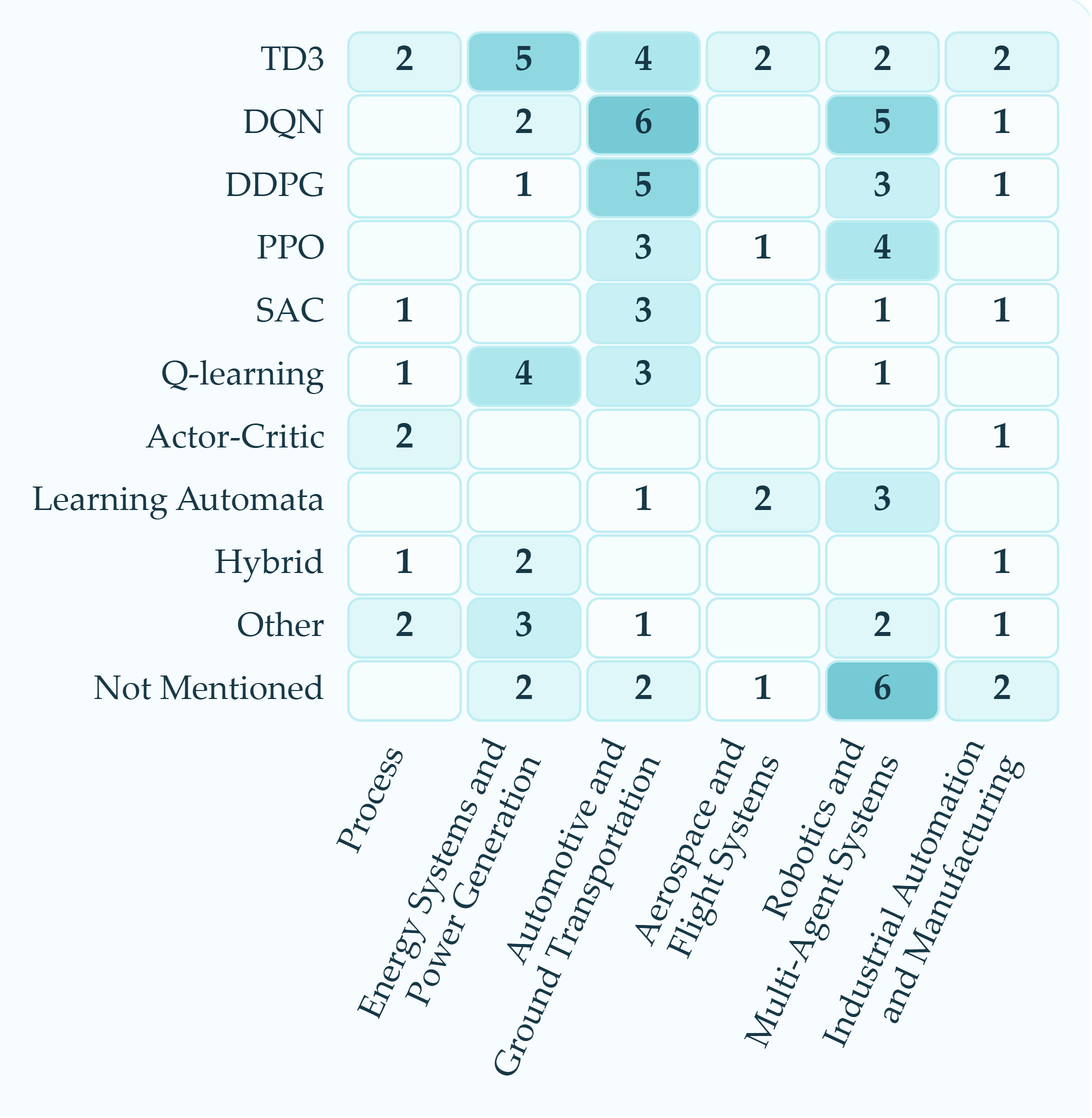}
	\caption{RL algorithms vs. application fields in the integration of MPC--RL for linear control}
	\label{Tax2D_RL_Type_Applications}
\end{figure}

\begin{itemize}
\item \textbf{Algorithmic and Functional Synergy:}
The pairing of specific MPC types with particular RL algorithms is heavily influenced by the fundamental role the RL agent is expected to perform. Figure \ref{Tax2D_MPC_RL_Type} illustrates the co-occurrence between MPC variations and the utilized RL algorithms (e.g., Value-based vs. Policy Gradient). Similarly, Figure \ref{Tax2D_MPC_RL_Role} details how the specific structural form of the MPC defines the boundaries of the RL agent's intervention--whether it acts as a dynamic tuner, a cost approximator, or a high-level reference generator. The intrinsic relationship between the chosen RL algorithm and its designated functional role within the control loop is further mapped in Figure \ref{Tax2D_RL_Role_RL_Type}.

\item \textbf{Application-Driven Methodological Choices:}
Control design is ultimately dictated by the physical and operational constraints of the application domain. As demonstrated in Figure \ref{Tax2D_RL_Role_Applications}, specific application fields exhibit strong preferences for certain RL roles; for instance, complex process systems often rely on RL for parameter tuning, whereas autonomous navigation heavily favors RL for high-level reference tracking. Furthermore, Figure \ref{Tax2D_MPC_Applications} maps the preferred MPC structures across various industries, while Figure \ref{Tax2D_RL_Type_Applications} outlines which RL algorithms have historically succeeded in overcoming the unique challenges of each physical domain.

\item \textbf{The Pivotal Role of Cost Functions:}
In the hybrid RL--MPC paradigm, the objective function serves as the critical mathematical bridge translating high-level operational goals into solvable optimization geometries. Figure \ref{Tax2D_Cost_Applications} demonstrates how different application fields mandate specific cost function forms (e.g., economic costs for energy systems versus quadratic penalties for precise robotic tracking). The dependency of these cost structures on the underlying MPC formulation is illustrated in Figure \ref{Tax2D_Cost_MPC}. Finally, Figure \ref{Tax2D_Cost_RL_Types} and Figure \ref{Tax2D_Cost_RL_Roles} provide a comprehensive overview of how cost function designs intertwine with both the RL algorithmic structure and the agent's functional role, dictating the ultimate learning efficiency and closed-loop stability of the integrated system.
\end{itemize}


\section{Conclusion and Future Research Directions}
\label{sec:conclusion}

This paper has presented a SLR on the integration of RL and MPC, with a specific focus on linear control systems. Through a rigorous methodological protocol, we synthesized the expanding body of literature to demystify the structural and algorithmic synergies that define this hybrid control paradigm. 

The core contribution of this review is the development of a comprehensive, multi-dimensional taxonomy. By systematically categorizing the literature, we identified that the efficacy of an RL--MPC architecture heavily depends on the precise functional role assigned to the RL agent--whether acting as a high-level reference generator, a dynamic parameter tuner, or a terminal cost approximator. Furthermore, our cross-dimensional analysis illuminated how specific application domains dictate the mathematical formulation of objective functions, as well as the selection of underlying MPC variants and RL algorithms. The insights drawn from these intersections provide a vital roadmap, enabling control engineers to select optimal hybrid architectures tailored to domain-specific constraints and operational objectives.

Despite the substantial progress observed in recent years, several critical challenges remain unresolved, paving the way for focused future research. Foremost among these is the urgent need for rigorous theoretical guarantees. While empirical successes of learning-based predictive control are abundant, establishing mathematical proofs for closed-loop stability, recursive feasibility, and robustness remains a formidable challenge. Future research must endeavor to bridge the theoretical gap between Safe RL formulations and traditional robust or stochastic MPC paradigms. Concurrently, computational complexity and real-time feasibility present significant bottlenecks. The simultaneous deployment of deep neural networks--acting as RL policies or value functions--alongside online optimization solvers imposes profound computational burdens. To alleviate latency issues, particularly in systems with fast dynamics, future investigations should explore the integration of explicit MPC with lightweight RL models or leverage advanced hardware solutions such as neuromorphic computing architectures.

Beyond theoretical and computational constraints, practical deployment hurdles such as sample efficiency and the sim-to-real gap demand targeted attention. Given that most RL agents require extensive interaction with high-fidelity simulators prior to physical deployment, enhancing sample efficiency through model-based RL, meta-learning, or offline RL is imperative. Mitigating the reality gap during hardware implementation is a crucial step for the industrial scalability of RL--MPC systems. Moreover, the systematic advancement of this field is currently hindered by the absence of standardized benchmarking. The literature urgently requires unified, open-source testbeds across diverse application domains, including robotics, energy systems, and process control. The development of such standardized frameworks will significantly accelerate comparative studies and ensure reliable methodological validation across the research community.

In conclusion, the reviewed literature indicates that the fusion of RL and MPC has become an important direction in modern control engineering, particularly for architectures based on linear or linearized predictive models.
 By intrinsically combining the rigorous safety boundaries of predictive control with the adaptive intelligence of machine learning, this synergistic architecture holds the transformative potential to solve some of the most intractable decision-making problems in complex, highly uncertain environments.

\bibliographystyle{elsarticle-num}
\bibliography{60_selected_7}

\end{document}